# The ACROBAT 2022 Challenge: Automatic Registration Of Breast Cancer Tissue

Philippe Weitz[*,1], Masi Valkonen[*,2], Leslie Solorzano[*,1], Circe Carr[2], Kimmo Kartasalo[1], Constance Boissin[1], Sonja Koivukoski[3], Aino Kuusela[2], Dusan Rasic[4], Yanbo Feng[1], Sandra Sinius Pouplier[4], Abhinav Sharma[1], Kajsa Ledesma Eriksson[1], Stephanie Robertson[5], Christian Marzahl[6], Chandler D. Gatenbee[7], Alexander R.A. Anderson[7], Marek Wodzinski[8,9], Artur Jurgas[8,9], Niccolò Marini[8,10], Manfredo Atzori[8,11], Henning Müller[8,12], Daniel Budelmann[13], Nick Weiss[13], Stefan Heldmann[13], Johannes Lotz[13], Jelmer M. Wolterink[14], Bruno De Santi[15], Abhijeet Patil[16], Amit Sethi[16], Satoshi Kondo[17], Satoshi Kasai[18], Kousuke Hirasawa[19], Mahtab Farrokh[20], Neeraj Kumar[20], Russell Greiner[20,21], Leena Latonen[3], Anne-Vibeke Laenkholm[4], Johan Hartman[5,22], Pekka Ruusuvuori[§,2,23], Mattias Rantalainen[§,1,22]

* contributed equally
§ contributed equally

1) Department of Medical Epidemiology and Biostatistics, Karolinska Insitutet, Stockholm, Sweden
2) Institute of Biomedicine, University of Turku, Turku, Finland
3) Institute of Biomedicine, University of Eastern Finland, Kuopio, Finland
4) Department of Surgical Pathology, Zealand University Hospital, Roskilde, Denmark
5) Department of Oncology and Pathology, Karolinska Institutet, Stockholm, Sweden
6) Gestalt Diagnostics, Spokane, USA
7) Department of Integrated Mathematical Oncology, Moffitt Cancer Center, Tampa, USA
8) Informatics Institute, University of Applied Sciences Western Switzerland, Switzerland
9) Department of Measurement and Electronics, AGH University of Kraków, Poland
10) Department of Computer Science, University of Geneva, Geneva, Switzerland
11) Department of Neuroscience, University of Padova, Italy
12) Medical Faculty, University of Geneva, Switzerland
13) Fraunhofer Institute for Digital Medicine MEVIS, Lübeck, Germany
14) Department of Applied Mathematics, Technical Medical Centre, University of Twente, Enschede, The Netherlands
15) Multimodality Medical Imaging, Technical Medical Centre, University of Twente, Enschede, The Netherlands
16) Department of Electrical Engineering, Indian Institute of Technology, Bombay, India
17) Graduate School of Engineering, Muroran Institute of Technology, Hokkaido, Japan
18) Faculty of Medical Technology, Niigata University of Health and Welfare, Niigata, Japan
19) FORXAI Business Operations, Konica Minolta, Inc., Osaka, Japan
20) Department of Computing Science, University of Alberta, Edmonton, Alberta
21) Alberta Machine Intelligence Institute, Edmonton, Canada
22) MedTechLabs, BioClinicum, Karolinska University Hospital, Stockholm, Sweden
23) Faculty of Medicine and Health Technology, Tampere University, Tampere, Finland

## Abstract

The alignment of tissue between histopathological whole-slide-images (WSI) is crucial for research and clinical applications. Advances in computing, deep learning, and availability of large WSI datasets have revolutionised WSI analysis. Therefore, the current state-of-the-art in WSI registration is unclear. To address this, we conducted the ACROBAT challenge, based on the largest WSI registration dataset to date, including 4,212 WSIs from 1,152 breast cancer patients. The challenge objective was to align WSIs of tissue that was stained with routine diagnostic immunohistochemistry to its

H&E-stained counterpart. We compare the performance of eight WSI registration algorithms, including an investigation of the impact of different WSI properties and clinical covariates. We find that conceptually distinct WSI registration methods can lead to highly accurate registration performances and identify covariates that impact performances across methods. These results establish the current state-of-the-art in WSI registration and guide researchers in selecting and developing methods.

## Introduction

Computational pathology is likely to significantly impact current routine clinical workflows in pathology labs. Applications range from the automation of routine procedures such as cancer detection and Gleason grading of prostate biopsies[1–3] to the prediction of diagnostic information that pathologists can not obtain from visually inspecting tissue, including prognosis [4,5], treatment response [5] molecular subtypes[6,7] or gene expression[8–11]. The rapid advancements of these methods in recent years have been enabled by progress in computer vision and the advent of digital pathology. In digital pathology, glass slides with tissue samples are digitised using whole-slide scanners, often at a magnification of 400, referred to as 40X, resulting in whole-slide-images (WSI) with a gigapixel scale. Pathologists then assess WSIs on a screen instead of physical glass slides with a microscope. Currently, the vast majority of methods in computational pathology are limited to WSIs of tissue stained with haematoxylin and eosin (H&E)[12,13]. However, the analysis of immunohistochemically (IHC) stained tissue, e.g. for biomarker scoring, is an essential component of the diagnostic workflow. The combination of information from multiple stains has the potential to unlock both novel research and clinical applications. Examples in research are stain-guided learning[14–16], virtual staining[17–20], the analysis of multiplex stained histology[21,22], 3D reconstruction[23,24] and the transfer of annotations or segmentations between WSIs[25–27]. In the clinical setting, multi-stain information can aid in the identification of regions of interest (such as invasive cancer) during biomarker scoring, or the investigation of suspicious lesions at resection margins.

The combination of information from multiple WSIs requires the spatial alignment of corresponding tissue areas between WSIs, which is referred to as WSI registration. WSI registration is a particularly challenging registration task due to the gigapixel scale, differences between the appearances of differently stained tissue, changes in appearance, structure and morphology between tissue regions in non-consecutive sections and the introduction of artefacts, tears and deformations during processing of the micrometre-thin tissue sections. WSI registration is therefore an active field of research.

In order to establish the current state-of-the-art in WSI registration, we organised the ACROBAT (Automatic Registration of Breast Cancer Tissue, acrobat.grand-challenge.org) challenge. For this challenge, we published the currently largest publicly available data set of matched H&E and IHC WSIs, consisting of 4,212 WSIs in total[28] and generated over 54,000 landmark points with 13 annotators. All WSIs in the data set originate from tissue sections that were generated during routine diagnostic workflows at the time of initial diagnosis. An example of an H&E WSI and corresponding IHC WSIs is depicted in Figure 1. The objective of the challenge was to align tissue in the IHC WSIs to corresponding tissue in the H&E WSIs. WSI registration has previously been addressed in the ANHIR challenge[29]. While the ANHIR challenge made valuable contributions to the field of WSI registration, it was limited by the high quality of sections and WSIs, which is not representative of clinical material, as well as by the availability of both training and test data with 355 WSIs in total, albeit originating from a wide variety of organs and stains.

Here, we describe the results of the ACROBAT registration challenge, in which we assess the performances and limitations of eight WSI registration algorithms. We evaluated accuracy and robustness of each method and performed a detailed analysis of the impact of different clinical covariates, such as cancer grade and biomarker statuses, and the registered tissue types, which to our knowledge is the first analysis of this kind in the histopathology domain. We expect that the results of this challenge will clarify the current state-of-the-art of multi-stain WSI registration and provide evidence for the integration of the analysed methods into future research studies and clinical applications. The findings of this study can furthermore guide the development of WSI registration solutions that generalise between stains and tissue types and that can be applied to WSIs that originate from routine diagnostics.

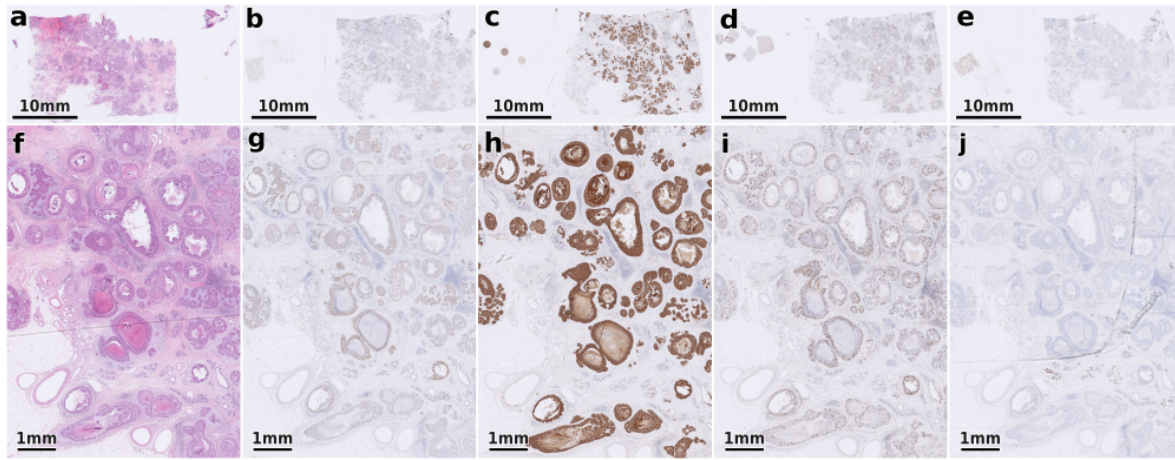

**Figure 1.** Example of an H&E stained tissue section and corresponding IHC stained tissue. The first row (a-e) shows an overview over the entire WSIs, whereas the second row (f-j) shows corresponding tissue regions at a higher magnification. a) and f) show the H&E WSI, b) and g) tissue stained with ER IHC, c) and h) with HER2 IHC, d) and i) with KI67 IHC and e) and j) with PGR IHC.

# Results

## Deep Learning has become an ubiquitously used tool for WSI registration that complements classical image analysis techniques.

A wide range of conceptually different WSI registration approaches were used in the challenge. Some methods relied on traditional image processing while others were based on deep learning. A common design pattern was to split the registration task into multiple steps, starting with image preprocessing, followed by an initial alignment and subsequently a deformable registration. The preprocessing step aims at normalising or discarding information e.g. by colour space transformation, contrast normalisation and downscaling to focus registration on essential information and simplify the problem for the subsequent steps. Teams VALIS, Fraunhofer MEVIS, and MeDAL target computations to meaningful areas through an initial tissue segmentation and then focus registration only on the detected tissue area. The initial alignment step roughly aligns images using translation, rotation, reflection, scaling or affine transformations in order to simplify the task for the remaining deformable step. The deformable step involves elastic or curvature-controlled transformations that attempt to completely align image contents and can account for complex deformations of the tissue. Registration steps were typically performed iteratively starting at low resolutions with increasing resolutions during subsequent steps. A common division of registration algorithms is into intensity-based and feature-based methods. Intensity-based methods rely on finding correspondences between image intensities based on a similarity function, whereas feature-based methods extract features, such as points, and extract a descriptor from their local neighbourhood to establish matching feature pairs between images.

Six out of the eight analysed teams used feature-based registration, with some relying on more recent approaches such as SuperPoint[30] and SuperGlue[31] while others used more classical approaches like SIFT[32], BRISK[33] and RANSAC[34]. For the intensity-based parts of participants' algorithms, the most common similarity criterion was cross-correlation (CC) and its variants normalised cross-correlation (NCC) and convolution. Only team Fraunhofer MEVIS employed normalised gradient fields (NGF)[35], which bases its matching on intensity gradient orientations. No participant proposed a solution utilising mutual information. Seven out of the eight top performing teams applied deep learning techniques in parts of their workflows. This shows deep learning has further permeated multi-stain WSI registration since the ANHIR challenge in 2019, where only one method applied deep learning. Table 1 summarises all the aforementioned aspects and underlines which parts of the groups' workflows involve deep learning, either during preprocessing or registration. For instance, Gestalt Diagnostics and AGHSSO used the graph neural network-based SuperGlue[31] for feature matching, and SK used the convolutional neural network-based registration framework Voxelmorph[36].

The use of external data was allowed in this challenge. The three teams Fraunhofer MEVIS, Gestalt Diagnostics and AGHSSO used external data for their algorithm development. Fraunhofer MEVIS trained a tissue segmentation model with external data, Gestalt Diagnostics used external data to optimise their final parametrization and AGHSSO further validated their model with data not provided by the organisers. More detailed summaries and additional information for each of the methods is available in the Supplementary Materials.

**Table 1 | Summary of methods**. A summary of each team's method with descriptions of their main workflow. More detailed descriptions are available in the Supplementary Materials. Underlined elements are where deep learning is used.

| Team | Code available | Uses deep learning | Initial alignment | Initial alignment criterion | deformable transformation | deformable criterion | Optimization | Preprocessing |
|---|---|---|---|---|---|---|---|---|
| Gestalt Diagnostics | | ✓ | Rotation search with Superglue/OpenGlue/LoFTR & RANSAC | No. of keypoint matches | Local affine for triangular partitions | RANSAC inliers | - | Contrast-normalising, random equalize, random sharpness, non-maximum suppression |
| VALIS | ✓ | ✓ | BRISK keypoints VGG descriptors & RANSAC | RANSAC inliers, Tukey inliers | DeepFlow[37] | DeepFlow's DeepMatch | - | Downsample, foreground segmentation, grayscale, invert intensities, intensity normalisation with Akima interpolation. |
| AGHSSO | ✓ | ✓ | SIFT/SuperPoint & RANSAC/SuperGlue | Sparse descriptor error | Multi-level and weighted local NCC optimization | NCC | Adam | Downsample, grayscale, invert intensities and equalize with CLAHE |
| Fraunhofer MEVIS | | ✓ | Align centers of mass + rotational search with NGF | NGF | Optimise control point grid with linear interpolation | NGF | L-BFGS | Downsample, grayscale, foreground segmentation |
| NEMESIS | ✓ | ✓ | SIFT & RANSAC + tissue mask overlap | RANSAC inliers + Dice score | Implicit neural representation of transform with MLP[38] | NCC | Adam | Downsample, grayscale, histogram equalisation, gaussian smoothing |
| MeDAL | ✓ | | Template matching (mask convolution, 1° intervals) | Convolution maximum | RBF in local keypoints found from tissue masks | convolution | - | Downsample, foreground segmentation, grayscale |
| SK | | ✓ | Template matching (NCC, 0° and 180° rotations) | NCC | Voxelmorph[36] | MSE | Adam | Downsample, grayscale, WSI border removal |
| MFRGNK | ✓ | | ORB & RANSAC with projective transform | RANSAC inliers | - | - | - | Downsample, Macenko colour normalization[39] |

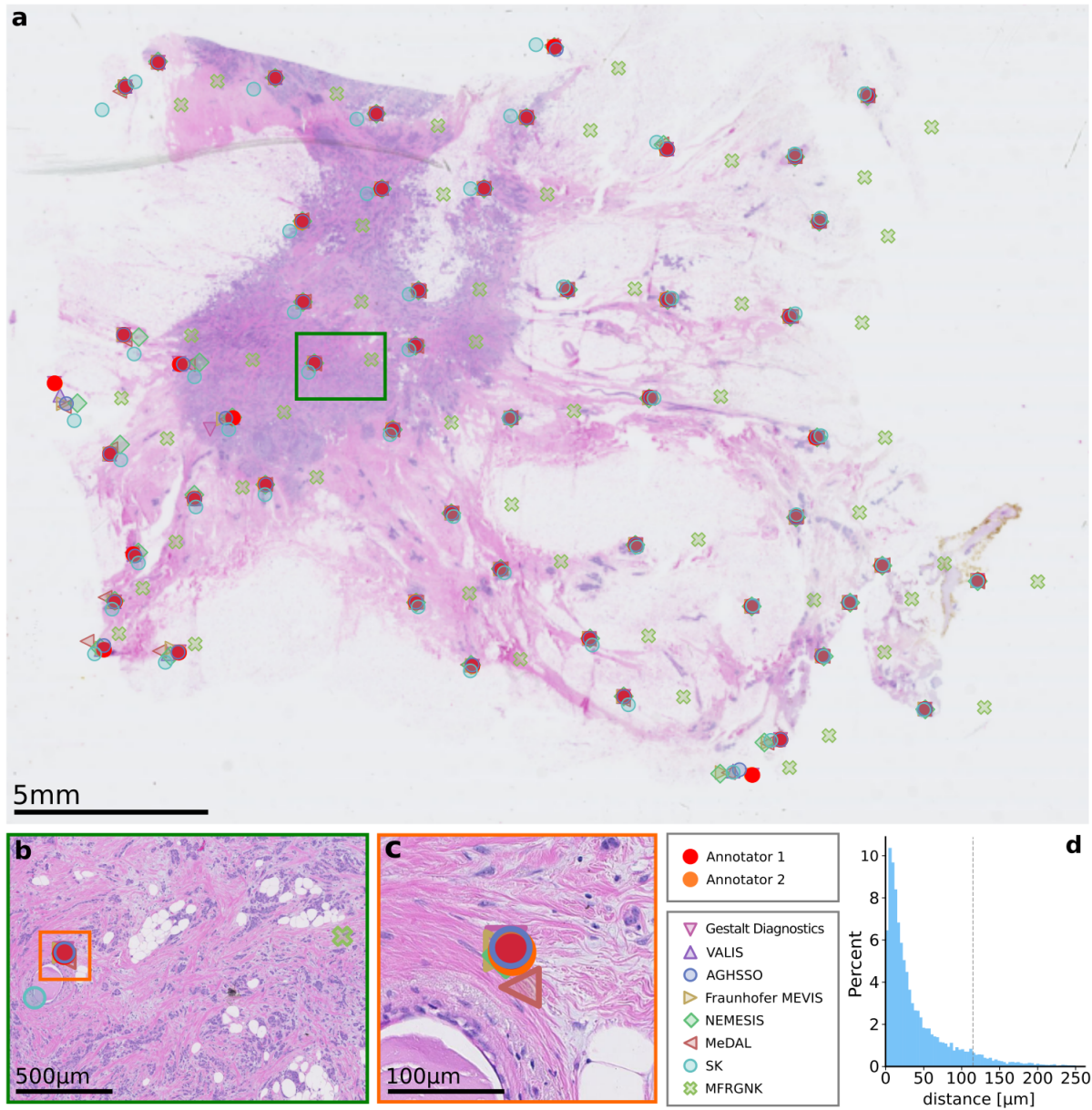

**Figure 2.** Example of a target H&E WSI and distribution of DBAs. a) shows an overview of a H&E WSIs with annotated and registered landmarks. b) and c) depict a closer view of a specific landmark. d) shows a histogram of the DBAs, where the dashed line indicates the DBA exclusion threshold for performance metric computation. The histogram is capped at 250 µm.

Diverse registration methodologies result in several performance clusters, one method approaches human annotator accuracy.

The primary ranking metric for the challenge was the median 90th percentile of the target registration error (TRE) within each WSI pair. To reduce the effect of human error, TREs were computed by averaging the distance in µm of registered landmarks to two target landmarks placed by two different annotators. An example of a target H&E WSI, with annotator and registered landmarks is depicted in Figure 2 a-c. As a quality control step, we excluded landmarks with poor agreement between annotators from performance metric computations based on distance between annotators (DBA) >115 µm, as indicated in the histogram of DBAs in Figure 2d. In total, 13,130 pairs of landmarks in 297 image pairs were included for metric computation. The distributions of the 90th percentiles in the validation and test data are depicted in Figure 3a. Figure 3b depicts scatterplots of 90th percentiles against the 90th percentiles of DBAs. Based on the primary ranking metric, the algorithm developed by Gestalt Diagnostics achieved the best score, with a median 90th percentile of 60.1 [55.8, 68.6] µm. This is approximately half the median 90th percentile of the methods that follow in the ranking, starting with VALIS with 123.3 [98.5, 144.1] µm, AGHSSO with 137.6 [120.3, 176.7] µm and Fraunhofer MEVIS with 155.3 [123.1, 184.7] µm. The solutions of NEMESIS and MeDAL are in the range of three to four times the lowest median 90th percentile, with 200.5 [176.7, 257.1] µm and 262.5 [225.4, 322.5] µm respectively. The median 90th percentile of SK of 1,230.0 [1,141.0, 1,341.5] µm is one order of magnitude higher and the solution of MFRGNK one order of magnitude higher compared to SK with 15,938.0 [15,117.0, 16,598.6] µm. A comparison of the distributions of the 90th percentiles in Figure 3a with two-sided Mann-Whitney U rank tests indicates that for Gestalt Diagnostics, 90th percentiles of TRE differ between the validation and test set, with Benjamini-Hochberg (BH) adjusted p-value < 0.01. For the other methods, this comparison reveals no differences. Both Figure 3a and b indicate that for all methods except Gestalt Diagnostic, there are outlier image pairs with considerably higher 90th percentiles of TRE. E.g. for VALIS, there is a higher number of outlier image pairs with poor registration quality compared to AGHSSO, which the median 90th percentile is robust against, but not the mean. Correspondingly, Figure 3c shows the ranking for each metric that is available in Table 2. The rankings are mostly stable across metrics. Only the algorithm proposed by VALIS is ranked lower compared to AGHSSO regarding the mean 90th percentile and the median error distance across all landmarks, as well as the mean distance reduction and AGHSSO and Fraunhofer MEVIS regarding the mean error distance across all landmarks. Supplementary Figure 2 shows the stability of the ranking with the median 90th percentile for varying exclusion thresholds in µm for the DBA. Only for exclusion thresholds below 70 µm, the ranking between AGHSSO and Fraunhofer MEVIS begins to depend on the threshold. Paired two-sided Wilcoxon signed rank tests indicate that the distributions of 90th percentiles are different between all submissions (compare Supplementary Figure 3) with BH-adjusted p-values < 0.01 for each comparison. Supplementary Figure 4a shows the Spearman correlations of median 90th percentiles in the test set, which reveals a cluster of correlations for the six top-performing methods with correlations ranging from 0.61 to 0.93.

    All median 90th percentiles, along with the 90th percentiles of 90th percentiles, mean 90th percentiles and the median and mean across all landmarks without WSI-wise aggregation are listed in Table 2. It also contains the slide-wise aggregated mean reduction in distance between source and target landmarks in percent, which may guide intuition on algorithm performances.

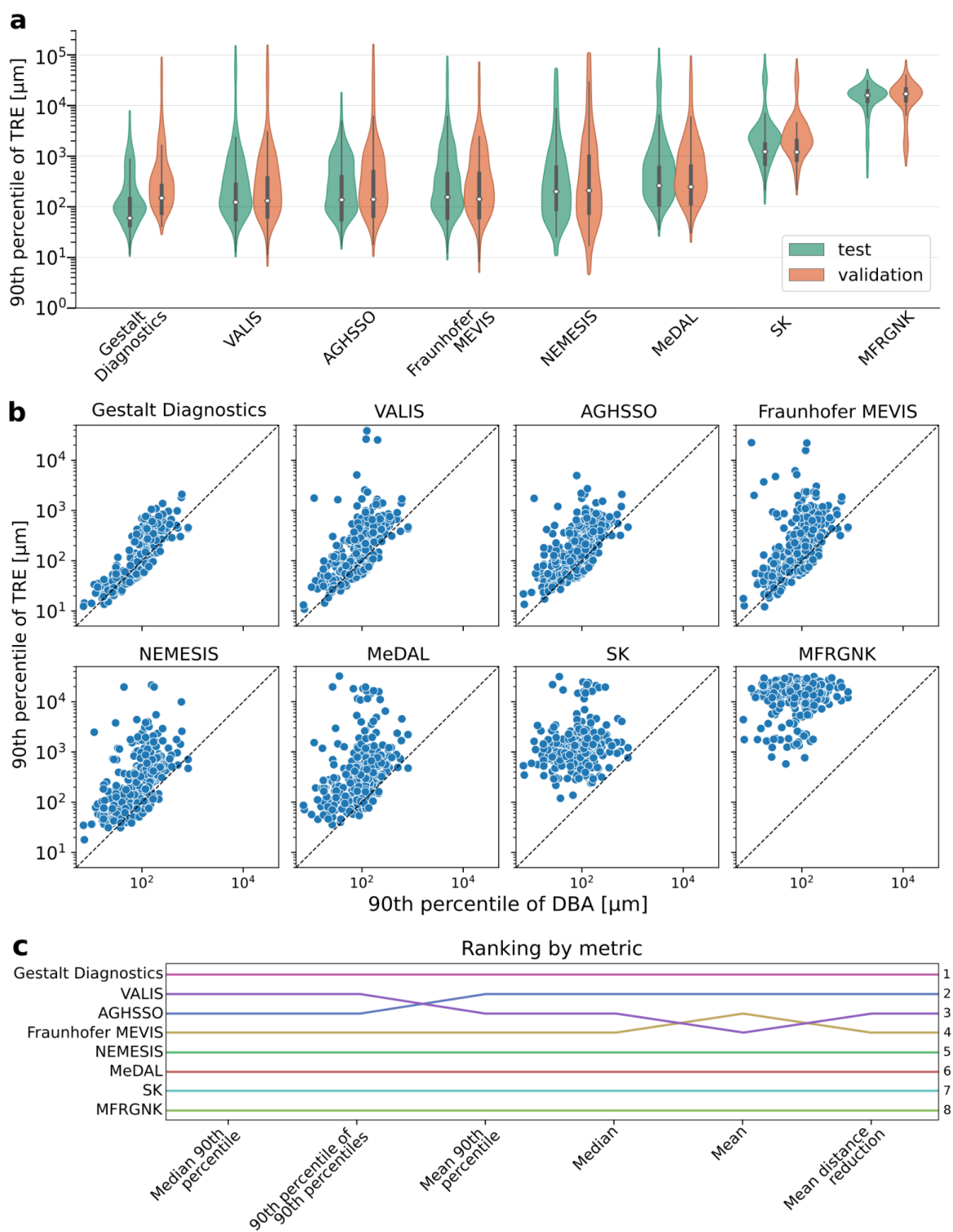

**Figure 3.** Overview of 90th percentiles of TRE and rankings. a) shows violin plots of the distributions of 90th percentiles of TREs in the validation and test data for all eight investigated methods, while b) shows scatterplots of the WSI-wise 90th percentiles of DBAs compared to the corresponding 90th percentiles of TREs, excluding landmarks with a DBA > 1mm when computing 90th percentiles for annotators. c) shows the ranking of methods for each of the metrics that are listed in Table 2.

In order to contextualise algorithm performances, we also computed all metrics for the DBA. The median 90th percentile of DBAs is 67.0 [62.2, 72.4] µm and therefore slightly higher compared to the value of Gestalt Diagnostics of 60.1 [55.8, 68.6] µm. It is possible to achieve a lower TRE than DBA by registering a landmark to a location between the two landmark positions chosen by the two annotators, as shown in Supplementary Figure 1. The mean 90th percentile of DBAs of 63.5 [60.3, 66.6] µm however is lower than the lowest corresponding TRE by Gestalt Diagnostics of 160.0 [134.0, 189.4], which is also the case for the mean across all landmarks of 31.1 [30.6, 31.6] µm, compared to 63.3 [60.1, 66.6] µm for the best-performing algorithm.

**Table 2.** *Metric values for the primary challenge metric, the median 90th percentile of error distances across WSIs, alongside further metrics that could be used to rank algorithm performances. Median and mean were computed on the landmark-level, without previous aggregation on the WSI-level. The mean distance reduction indicates the mean reduction in distance between source and target landmark position due to the registration. Confidence intervals were obtained by bootstrapping with 10,000 bootstrap samples.*

| Team | Median 90th percentile [µm] | 90th percentile of 90th percentiles [µm] | Mean 90th percentile [µm] | Median [µm] | Mean [µm] | Mean distance reduction [%] |
|---|---|---|---|---|---|---|
| Gestalt Diagnostics | 60.1 [55.8, 68.55] | 449.64 [345.97, 535.96] | 159.99 [132.78, 188.43] | 22.29 [21.8, 22.72] | 63.29 [60.05, 66.59] | 98.97 [98.73, 99.18] |
| Annotators | 66.99 [62.19, 72.37] | 97.72 [95.26, 101.44] | 63.47 [60.34, 66.66] | 21.27 [20.79, 21.8] | 31.09 [30.61, 31.56] | n.a. |
| VALIS | 123.32 [98.49, 144.12] | 694.45 [580.41, 857.44] | 578.93 [274.38, 966.9] | 37.98 [37.01, 38.99] | 313.36 [275.43, 353.76] | 97.41 [96.21, 98.33] |
| AGHSSO | 137.63 [120.9, 175.65] | 713.4 [604.16, 838.94] | 303.16 [256.04, 358.43] | 34.64 [33.75, 35.46] | 122.21 [117.03, 127.63] | 98.2 [97.86, 98.5] |
| Fraunhofer MEVIS | 155.29 [123.1, 184.65] | 1019.5 [856.2, 1343.29] | 604.35 [393.19, 870.49] | 40.88 [39.93, 41.93] | 294.14 [269.65, 319.55] | 96.56 [95.37, 97.59] |
| NEMESIS | 200.47 [176.69, 257.13] | 1308.64 [1013.13, 1834.37] | 733.04 [510.86, 1010.42] | 62.72 [60.96, 64.26] | 349.89 [325.62, 375.91] | 95.79 [94.37, 97.0] |
| MeDAL | 262.49 [225.44, 322.47] | 1607.82 [1177.53, 2558.87] | 1221.55 [838.07, 1673.09] | 81.9 [79.63, 83.92] | 721.24 [674.45, 768.83] | 93.24 [91.18, 95.13] |
| SK | 1230.01 [1141.98, 1341.52] | 3292.55 [2539.85, 4722.35] | 2438.45 [1956.03, 2981.57] | 628.44 [612.02, 644.17] | 1524.35 [1466.42, 1582.3] | 84.3 [81.63, 86.83] |
| MFRGNK | 15938.02 [15117.95, 16576.21] | 22946.95 [21964.99, 23844.39] | 15342.27 [14576.8, 16107.18] | 9224.71 [9064.66, 9400.27] | 9988.07 [9876.3, 10101.04] | 29.23 [26.38, 32.07] |

We also investigated failure cases to identify how algorithms can be improved further. The image pair with the worst mean 90th percentile across methods is depicted in Supplementary Figure 5. The high degree of cropping does not allow for the detection of a tissue outline and therefore a reliable initial alignment for some methods. While this impacts algorithms significantly, with 90th percentiles of 440.90 µm for Gestalt Diagnostics, closely followed by MeDAL with 442.61 µm and NEMESIS with 586.73 µm, it is not challenging for human annotators to find corresponding landmarks, with a 90th percentile of distances between first and second annotator of 98.47 µm.

## Linear mixed effects model analysis reveals covariates that consistently impact TREs across algorithms.

In order to identify which properties of the landmarks and image pairs impact algorithm performances, we conducted a linear mixed effects (LME) model analysis for each team, with the *log*10-transformed landmark-wise TREs or DBAs as the endogenous variable. The analysis is adjusted for the slide ID and combination of first and second annotator as random effects. Percentage changes for one unit increase of the fixed effects for the LMEs for the TREs of the six best performing teams and the DBA are depicted in Figure 4a. Supplementary Figure 6 shows the corresponding fixed effects coefficients, whose values are available in Supplementary Table 4. Due to the *log*-transform, percentage changes of the respective reference TRE accumulate multiplicatively across effects. For covariates for which the 95% confidence intervals of most or all fixed effects include zero, it is nevertheless possible to observe and interpret trends.

As depicted in Figure 4a, the antibody of the stain of the IHC WSIs in the image pairs compared to ER as the reference category appears to not impact algorithm performances across methods, potentially with the exception of HER2, where the point estimates of the effect sizes are consistently below zero. For each landmark, a semantic segmentation of the surrounding tissue is available, including invasive cancer (IC), non-malignant changes (NMC), artefacts, ductal carcinoma in situ (DCIS), lobular carcinoma in situ (LCIS) and normal tissue as the reference class. With the exception of artefacts, landmarks in all segmentation classes are associated with a lower TRE compared to normal tissue. This effect is particularly pronounced for NMC, DCIS and IC. For landmarks within the IC regions, we also included the Nottingham histological grade (NHG) as an interaction. Both for NHG 2 and 3, which have less clearly defined growth patterns than NHG 1, the registration error increases across methods. Besides the NHG, we also modelled an interaction between the IC region, the biomarker status (BS) as assigned at time of diagnosis and the IHC stain. It appears that there is a trend towards higher TREs for landmarks within HER2-positive IC regions and potentially a weaker trend towards higher TREs in PGR-positive IC regions. KI67 and ER BS within IC regions are not associated with TRE.

Besides categorical fixed effects, we also analysed two continuous effects, the slide age and the distance of a landmark to the centre of tissue mass. The increase in error in percent with increasing units is depicted in Supplementary Figure 7. The slide age is strongly associated with the TRE. With the exception of Gestalt Diagnostics, all teams have an increase of TRE compared to the respective reference of approximately 100% at four years. For NEMESIS and MeDAL, TRE is also relevantly associated with the distance to the centre of tissue mass with an increase of 60% at 10 mm, at which there is an increase of ca. 20% for VALIS, AGHSSO and Fraunhofer MEVIS. In contrast, there is a weak negative association for Gestalt Diagnostics and the annotators.

Figure 4b shows the distributions of changes in percent for the estimated conditional means for the random effects. The interquartile range for the annotator combination is highest for the DBA, followed by Gestalt Diagnostics and decreasing with decreasing ranking, whereas the interquartile range for the slide ID is lowest for the annotators, followed by Gestalt Diagnostics and roughly increasing with decreased ranking. The correlations between conditional means for the slide ID are shown in Supplementary Figure 4b and closely resemble the correlations of the 90th percentiles, with a cluster for the six highest ranked methods but weaker correlations with the annotators.

Across fixed effects, the direction of statistically significant coefficients is the same throughout teams, with the exception of the distance to centre for Gestalt Diagnostics. The DBA is associated with fewer fixed effects than the registration methods, and effect sizes are generally smaller.

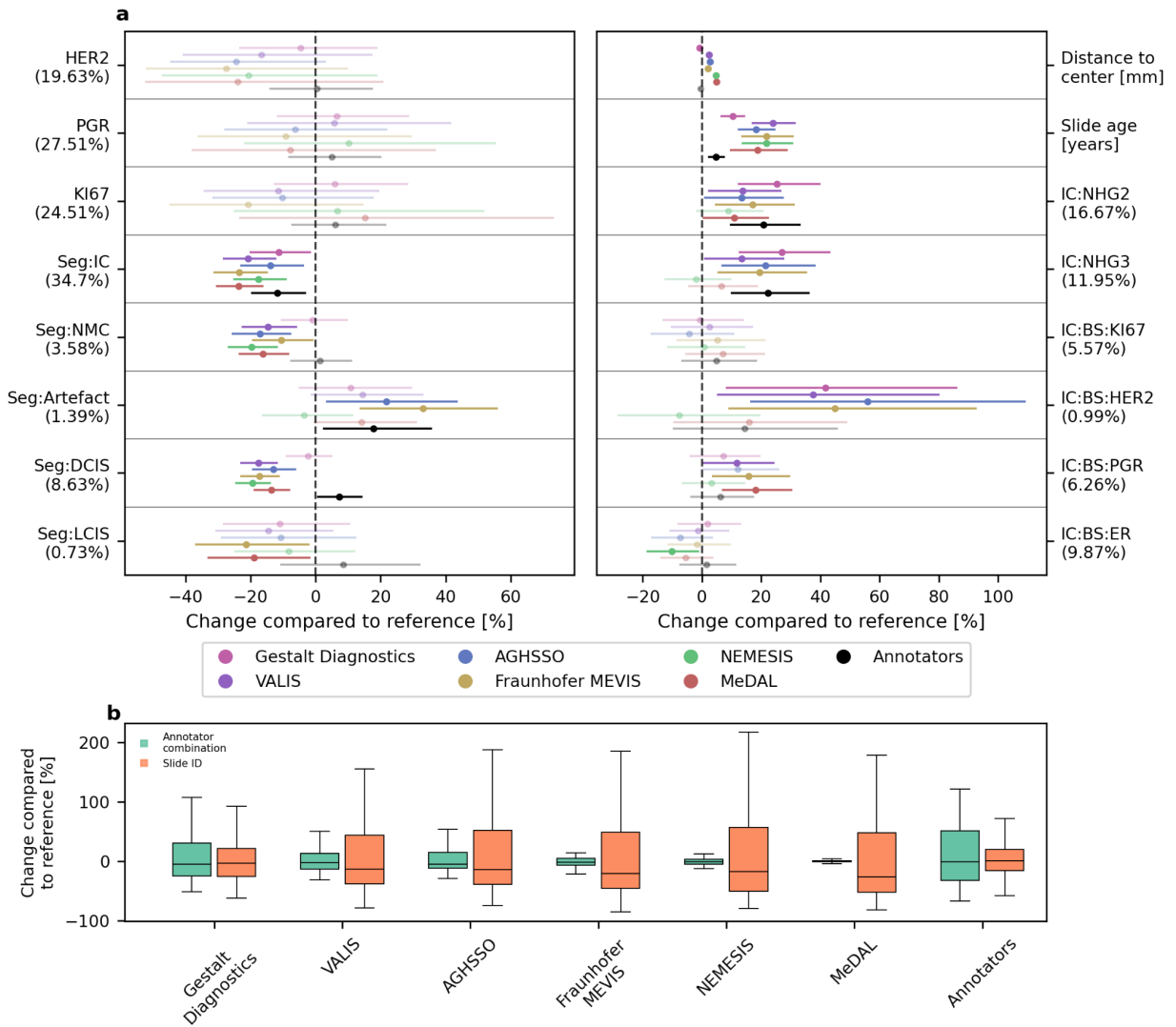

**Figure 4.** Coefficients and conditional means of random effects of the LME analysis for the TREs of the six highest ranked teams and the DBA for annotators. a) shows the percentage changes in TRE or DBA for one unit increase of fixed effects coefficients with 95% confidence intervals. Transparency of the respective marker is increased if the confidence interval includes zero. For categorical fixed effects, there are indications of the percentage of landmarks that are part of the respective category. For continuous effects, the unit is indicated. Effects starting with *Seg* indicate landmark tissue classes, with normal tissue as the reference class. b) shows boxplots that represent the distributions of the percentage changes of estimated conditional means of the random effects for the annotator combination and slide IDs. Boxes include the lower to upper quartile of data. Whiskers extend 1.5 times the interquartile range from the box outlines or the minimum or maximum value. Outliers outside of this range are not shown but are available in Supplementary Figure 6.

## Discussion

We organised the ACROBAT challenge to determine the state-of-the-art of multi-stain WSI registration algorithms and to test the applicability of current solutions for a real-world data set with slides from clinical routine. We published the largest-to-date data set for histology image registration, placed over 54,000 landmark points with 13 annotators for performance quantification and conducted an in-depth analysis of registration methods including clinical information and tissue segmentations.

A wide range of conceptually different approaches were evaluated in the challenge, showing that since the ANHIR challenge, deep learning has become an impactful tool also in the WSI registration domain. The best performing method was by Gestalt Diagnostics in terms of accuracy and robustness. In this method, a tree structure of triangular partitions is constructed using DL-based feature matching. Most top performing methods relied upon feature-based registration, which shows its effectiveness with challenging data sets. Also the use of external data may have played a role in the outcome as its utilisation increased at the better ranks. The obtained results suggest that feature-based approaches are able to achieve higher robustness than alternative methods, especially those using modern DL-based methods such as SuperPoint and SuperGlue. Feature-based methods may be less impacted by certain imperfections in the data such as tears. They have the advantage of being independent of foreground tissue segmentation and therefore make no assumptions of the presence of background. Intensity-based registration with cross-correlation is able to produce good results with grayscale converted H&E and IHC images when compared to more advanced multi-modal similarity metrics such as the NGF. One possible explanation for the high performance of the method by Gestalt Diagnostics is its fully feature-based approach. Adjacent histological sections have structures that are shared between them but also smaller structures that are not shared such as some of the cells, which can appear only on one of the sections. Feature-based registration may be able to focus on those shared structures between the images through the feature matching step. Another possible explanation is the lack of regularisation in the applied transformation. Other teams among the four best performing methods regularised the transformation to produce smooth deformations, which however comes at the cost of accuracy. While registration without regularisation can produce smaller errors in a setting with landmark based evaluation, there can be practical reasons to trade off accuracy for smoother transformations, e.g. if downstream analyses of the registered images require less distorted tissue.

Given the challenging nature of the data set with slides originating from routine clinical workflows, top-performing methods should be considered to have a high performance, both with regards to accuracy, as well as to robustness. Breast cancer cell diameters extend approximately up to 20 μm. The lowest mean TREs across all landmarks of 63.29 μm and 122.21 μm therefore cannot be assumed to allow a cell-level registration, but neighbourhoods of cells can be assumed to be registered correctly. Furthermore, depending on the section spacing, actual cell-level correspondence between the sections is impossible to determine. Therefore, the performance level achieved by the best-performing methods may already have reached the limit set by the technical setup using non-consecutive sections. While there are differences in performance based on the computed metrics, the mean distance reductions in percentage provide an intuition that all top-performing methods are well suited to significantly reduce the initial TRE. The ranking of methods is mostly stable both across metrics and annotator disagreement exclusion thresholds for landmarks. Statistical testing indicates that the ranking based on the 90th percentiles is unlikely to arise by chance. Nevertheless, methods with a lower ranking could be shown to be capable of similar performances through better algorithm optimization in future work. The rankings in this study however provide a clear indication of which methods are currently preferable. A direct comparison of the registration performances to the ANHIR challenge proves difficult, since TREs were normalised with image diagonals in the ANHIR challenge, rather than provided in μm.

Correlations among 90th percentiles of TREs are notably higher among algorithms than between algorithms and corresponding 90th percentiles of DBAs, with the exception of Gestalt Diagnostic. This finding is also supported by the almost identical correlations of the conditional means of the random effect that captures the slide ID. This indicates that image pairs that were difficult to annotate were not necessarily the same as those that were difficult to register for the proposed methods, while generally the same image pairs were difficult to register for the six top-performing methods. The image pair with the worst registration performance based on the 90th percentiles emphasises the importance of the initial alignment step, which fails in this case due to the precropping of the IHC WSI. This indicates that it would be worthwhile to focus future work in WSI registration on increased robustness against comparable failure cases.

In order to investigate which properties of the WSIs and tissue impact algorithm performances, we conducted LME analysis. The conditional means of the random effects of slide ID and annotator combination indicate that these effects have a higher impact on the TREs than the fixed effects. An analysis of the fixed effects shows that the antibody of the IHC stain within the image pair does not have a significant impact on the TREs, potentially with the exception of a trend towards lower TREs for HER2 WSIs compared to ER WSIs. This might be because the appearances of the routine diagnostic stains in breast cancer are relatively similar. Landmarks in HER2 BS positive IC regions are associated with higher TREs in IC regions, but BS seems to otherwise not impact TREs. It was not possible to investigate the BS in DCIS and LCIS regions, since BS is not routinely reported there, yet cells within these regions can be largely positively stained. Future research should investigate how visually more different stains impact TRE. Nevertheless, this means that the risk for biases due to IHC and BS in multi-stain studies that involve registration might not be high. Regions of DCIS, LCIS, and NMC both were over-annotated by annotators and appear to be associated with lower TREs compared to normal tissue. The reason for this might be the presence of more visually easily distinguishable structures, which could also explain the lower TRE in IC regions. Nevertheless, we find the reduced TRE in IC regions surprising, since IC is characterised through diffuse growth patterns. Potentially, increased nuclear density in IC regions leads to higher contrast reference points, which could be beneficial at lower resolutions. IC is likely to be located in the centre of resections, but the negative association with the TRE remained when adjusting for the distance to the centre of tissue mass. Higher NHGs are associated with increases of TRE and DBA in IC regions, which is likely due to increasingly poor differentiation of structures and cells. This indicates that there is some risk of bias from cancer grades in studies that deploy registration. The distance of landmarks to the centre of tissue mass is positively associated with TREs for most methods, indicating room for improvements in deformable registration. The covariate with the highest impact on the TRE across algorithms is the slide age, which can more than double the baseline registration error for the oldest slides for some methods. In registration-based studies that focus on outcomes, the slide age could therefore confound analyses in multiple ways. However, this is only a concern for studies that focus on WSIs from archived tissue sections. During deployment, WSIs would be generated from recently sectioned tissue.

While it was not possible to investigate the section spacing in this study due to lack of recorded information, the LME analysis adjusts for this through the random effect that captures the slide IDs, alongside other slide-specific properties. The conclusions of the LME analysis regarding the fixed effects are therefore likely transferable between section spacings, nevertheless, future research that elucidates the effect of section spacing would be of high interest. Interestingly, the interquartile range of the conditional means of the random effect that captures the annotator combination are highest for Gestalt Diagnostics. This observation is in concordance with the lowest slope in the reduction of the median 90th percentile of TRE in Supplementary Figure 2a and the significant improvement between validation and test set performance for Gestalt Diagnostics. This indicates that the performance estimation for Gestalt Diagnostics could be significantly limited through uncertainty in the annotations. While this also impacts the other teams, the relative impact on their performance metrics might not be as relevant and only the ranking between Fraunhofer MEVIS and AGHSSO could change depending on the DBA exclusion threshold. While the precision of landmarks might be a limiting factor for the quantification of algorithm performances, the LME analysis indicates that the DBA is only weakly associated with the covariates, which means that the TRE quantification is likely not biased within covariate categories.

Besides the precision of landmark placing, this study has several further limitations. It is important to consider that landmarks only allow the quantification of registration performances in sparsely sampled locations. While we hope that the performance in these points is a reliable proxy for the registration performance in all image regions, this is not guaranteed. Annotators over-annotated the classes DCIS, LCIS, NMC and LI and under-annotated normal tissue and artefacts. While this biases the evaluation of performance metrics towards the performance within the over-annotated classes, the correct registration of these tissue classes compared to normal tissue is likely of higher interest for future applications. We therefore think that this does not adversely impact the generalizability of quantitative results. Our evaluation is purely quantitative, no qualitative analysis was performed. Therefore, there is no guarantee that transformed images are feasible for possible downstream analyses following registration. The LME analysis is limited by the assumptions of additive relationships between covariates and a linear relationship between the covariates and the *log*10-transformed TRE. Furthermore, the high correlation between the WSI scanner, slide age and presence of control tissue does not allow to conclusively disentangle these effects. Another limitation of this challenge is the evaluation of registration algorithms as a whole. It could yield valuable insights to analyse different combinations of suggested pre-processing, initial alignment and deformable registration methods. Furthermore, we did not assess computational performances, since registration took place on the participants computing infrastructures.

Despite its limitations, we think that the ACROBAT 2022 challenge has elucidated the state of multi-stain WSI registration algorithms and their application to real-world data that originates from routine clinical workflows. While WSI registration is not yet a solved problem, the results in this study indicate that it has now become a sufficiently reliable technology to enable novel areas of research. Clinical applications are likely also possible with the observed registration performances, but would require further validation to ensure patient safety. Furthermore, this study has led to novel insights into specific strengths and weaknesses of current WSI registration methods and the mixed effects models analysis could be a model for future analyses of registration methods also outside of computational pathology. Five of the discussed methods are available under open-source licences, and we believe that this study has generated sufficient evidence of algorithm performance and robustness to warrant the proliferation of top-performing methods into a wide range of future applications. The ACROBAT 2023 challenge will build on the results of the ACROBAT 2022 challenge and investigate computational performances and algorithm performance under domain shifts.

# Methods

## Challenge Design

The ACROBAT challenge took place between April and September 2022. The objective of the challenge was the fully automatic registration of test set landmarks that were provided for the IHC WSIs to their H&E counterparts. For the training data, no landmarks were available, methods that were optimised with the training data therefore needed to be optimised in an unsupervised manner. Both for the validation and test data, IHC landmarks were published. Registered validation set landmarks could be submitted up to two times daily on the challenge website (acrobat.grand-challenge.org) in order to receive automated feedback on the algorithm performance. Test set performance was computed by the challenge organisers after the end of the challenge timeframe and no feedback on algorithm performance in the test set was available before the challenge workshop, which was held in conjunction with the MICCAI (Medical Image Computing and Computer Assisted Intervention) 2022 conference in Singapore. During the challenge, there were 221 submissions by 16 teams for the validation data. Eight methods qualified to be evaluated in the test set by submitting test set landmarks and an algorithm description before the challenge deadline. These eight methods are assessed in this publication.

## Data set

The ACROBAT data set consists of 4,212 WSIs from 1,153 female patients with primary breast cancer. All WSIs of a case contain tissue from the same tumour block, but sections are not necessarily consecutive. The cases were divided into a training set consisting of 750 cases (3,406 WSIs), a validation set consisting of 100 cases (200 WSIs) and a test set consisting of 303 cases (606 WSIs). Clinical characteristics of the test set cases are available in Supplementary Table 1. For each case in the training set, one H&E WSI and up to four IHC WSIs from the routine diagnostic stains ER, PGR, HER2 and KI67 are available. In the validation and test set, there is one H&E WSI and one randomly selected IHC WSI out of the four IHC stains available. All WSIs were digitised on either one of two NanoZoomer XRs or a NanoZoomer S360 at ca. 0.23 µm/pixel. The ACROBAT data set was published as pyramidal TIFF WSIs with a resolution of 0.92 µm/pixel at the highest magnification. Further details of the data generation and processing workflows are available in the ACROBAT data set descriptors[28,40]. Figure 1 shows an example of an H&E WSI from the data set with corresponding IHC WSIs.

## Landmark Annotations

Registration performance in the ACROBAT challenge was quantified based on landmark annotations. Annotations were generated by 13 members of the ABCAP research consortium (abcap.org), all of which have received histopathology education and who have previously worked with WSIs in research projects. Two of the annotators have pathologist training. Annotations were generated using a version of TissUUMaps[41] that was customised for the ACROBAT challenge. All annotations were generated using WSIs with 40X magnification as the highest resolution. For the validation data, annotations for each image pair were generated by a single annotator, whereas in the test data, each image pair was annotated by two annotators. Annotations were generated in two phases, the first of which was applied to both the validation and test data, whereas the second was only applied to the test data. Annotation protocols for both phases are available online.

During the first phase, annotators were asked to place 50 corresponding landmarks in an H&E-IHC image pair that was displayed side-by-side in TissUUMaps, placing first the IHC and then the H&E landmark. For the second phase, landmarks in the IHC WSIs were fixed in place and displayed, whereas landmarks in the H&E WSIs were randomly moved by ± 500 pixels, which corresponds to ±115 µm at 40X magnification. A second annotator was then selected randomly such that the first and second annotator were always different, with 87 combinations between annotators present in the data. The second annotator was then tasked with moving the H&E landmark to the locations that they considered to match the displayed IHC landmark. This results in 10,040 landmarks in the validation and 44,760 landmarks in the test set. In total, 13 annotators annotated 54,800 points with an average of 49.24 landmarks per image. Finding corresponding structures from the images was challenging for some image pairs due to the drastic difference in visual appearance and the number of annotations for those was often less than the set target of 50 point pairs per image pair. Together, from the test and validation sets 83% of the image pairs had exactly 50 landmarks, 6% less than 50 landmarks, and 11% above 50 landmarks. The majority of landmarks were placed in close proximity to each other by the first and second annotator, with 50% of the landmarks less than 20 µm apart and 80% less than 60 µm.

We then proceeded to exclude landmarks in the test set for which the distance between the location selected by the first and second annotator exceeds 115 µm to filter out annotations with low inter-annotator agreement. The distribution of distances between annotators is depicted in Figure 2 d). We furthermore excluded WSIs from evaluation for which fewer than 10 landmarks remained after this exclusion to ensure sufficient landmark density. This results in 13,130 landmarks and 297 WSIs included for final performance evaluation.

For 290 out of these 297 H&E WSIs in the test set, semantic annotations were generated by a trained pathologist who specialises in breast cancer. Annotations include the classes invasive cancer (IC), ductal carcinoma in situ (DCIS), lobular carcinoma in situ (LCIS), non-malignant changes (NMC), artefacts, lymphovascular invasion (LI) and tissue. We can therefore assign one of these classes to the majority of landmarks in the test set for analysis. Supplementary Table 2 lists the percentages of landmarks in each class, as well as the area percentage of each class in the total tissue area. Tissue and artefacts were under-annotated, with factors of 0.79 and 0.53. This could be explained by a lack of structures in tissue. Artefacts may only exist in one of the two images of an image pair. The proportion of IC landmarks corresponds to the proportion of IC area in the data set, with a factor of 1.03. DCIS, LCIS, NMC and LI were over-annotated, with factors of 4.73, 7.77, 2.88 and 9 respectively. For 3.64% of landmarks, the class assignment differs, which might e.g. be common for landmarks that were placed on edges of structures.

## Performance Evaluation & Ranking

Performance evaluation was based on the target registration error (TRE). For each registered landmark, there are two target landmarks by two different annotators. The H&E WSIs are the target images for the registration, whereas the IHC WSIs are the source images. The transformation that is found during the registration is therefore applied to the IHC landmarks, in order to transform their coordinates to the H&E coordinate system. For each registered IHC landmark, we computed the distance in micrometres to each of these two H&E target points and used the mean distance as the error distance, which we will refer to as TRE. Within each WSI, we aggregated these error distances into a WSI-level score by taking the 90th percentile of the distances. We chose the 90th percentile in order to emphasise robustness. In the case of missing landmarks in the submissions, we used the coordinates of the unregistered landmarks in the source image, capped at the image borders, to compute error distances. Submissions were then ranked on the median of the 90th percentiles. Submissions were ranked in two leaderboards, one including all eligible test set submissions, the other one only those submissions for which the code was made publicly available. Monetary prizes were then allocated to the first three teams in each leaderboard.

Beyond the median 90th percentile, we also computed the 90th percentiles of 90th percentiles, the mean 90th percentiles and the mean and median error distances across all landmarks without slide-level aggregation. Furthermore, we computed the mean reduction in the TRE in percent from unregistered landmark locations to the target locations.

## Linear Mixed Effects Model Analysis

In order to investigate the impact of different properties of individual landmarks on the resulting error distances, we fitted Linear Mixed Effects (LME) Models with the *R* package *lme4*[42]. One LME was fitted for each team and one for the annotators. A LME is a linear model of the form $y = X\beta + Zu + \epsilon$. Here, $y$ is a vector of the *log*10-transformed TREs for the teams and *log*10-transformed DBAs for the annotators for all landmarks in micrometres. The *log*-transform was necessary to ensure that residuals were approximately Gaussian. For the DBA LME model, we chose 1 mm as the exclusion threshold. $\beta$ represents the fixed effects coefficients and $u$ the random effects coefficients, with $X$ and $Z$ as matrices that contain the values of observations of covariates in their rows. $\epsilon$ denotes a random error term. Fixed effects covariates are independent of each other, whereas random effects are sampled from the same statistical units. Here, we consider each WSI as a statistical unit containing multiple landmarks, with 272 units for which all required information is available for inclusion in the LME analysis. Furthermore, the combination of first and second annotator is considered as a statistical unit, with 86 units in total. Both of these are therefore modelled as random effects in the LMEs. We included 16 fixed effects into the analysis. There are two continuous covariates, the slide age ranging from 0 to 5 years and the distance of landmarks from the centre of mass of the respective tissue mask in mm. HER2, PGR, KI67 indicate the IHC antibody used in the IHC WSI of an image pair, with ER as the reference category. With respect to semantic segmentation classes that landmarks are positioned in, landmarks can be assigned IC, artefact, DCIS, LCIS and NMC, with normal tissue as the reference category. We excluded landmarks with disagreeing tissue class between first and second annotator and LI landmarks, since there are too few of these to model. IC:NHG2 and IC:NHG3 indicate the grading of the invasive cancer region for the landmarks within the cancer region for an image pair, with IC:NHG1 as the reference category. IC:BS:KI67, IC:BS:HER2, IC:BS:PGR and IC:BS:ER indicate the clinical biomarker status (BS) of the respective antibody for landmarks within the invasive cancer region of a WSI. The biomarker status is considered as positive for ER and PGR above a threshold of 10% and 20% for KI67. HER2 was considered as negative based on IHC scores 0 to 1+, positive for IHC scores 3+ and negative or positive for 2+ depending on additional in situ hybridisation that assesses gene amplification. Biomarker statuses were assigned according to clinical guidelines at the time of diagnosis. Coefficient values whose 95% confidence interval does not include 0 will be considered as different from 0 and therefore as associated with the error distance. As shown in Supplementary Figure 9, the slide age, scanner model and presence of control tissue in the IHC WSI of an image pair are highly correlated. We therefore only included the slide age into the LME analysis to avoid collinearity. A scatterplot of the LME model residuals, a quantile-quantile plot and a plot of the density of residuals are available in the section of the Supplement that describes the algorithm of the respective team.

# Data Availability

The ACROBAT data set is available at https://snd.gu.se/en/catalogue/study/2022-190.

# Code Availability

Code used for displaying landmarks in a surrounding tissue region, code for computing registration performance metrics, as well as the annotator protocols are available from github.com/rantalainenGroup/ACROBAT.

## Acknowledgements

We acknowledge support from Stratipath and Karolinska Institutet sponsoring the ACROBAT challenge prizes; MICCAI society for hosting the ACROBAT challenge, and Nguyen Thuy Duong Tran for support with digitising histopathology slides.

We acknowledge funding from:
Vetenskapsrådet (Swedish Research Council)
Cancerfonden (Swedish Cancer Society)
ERA PerMed (ERAPERMED2019-224-ABCAP)
MedTechLabs
Swedish e-science Research Centre (SeRC)
VINNOVA
SweLife

Academy of Finland (#341967, #334782, #335976, #334774)
Cancer Foundation Finland
University of Turku Graduate School
Turku University Foundation

Oskar Huttunen Foundation
David and Astrid Hägelén Foundation
Orion Research Foundation
KI Research Foundation

This project has received funding from the Innovative Medicines Initiative 2 Joint

Undertaking under grant agreement No 945358. This Joint Undertaking receives support from the European Union's Horizon 2020 research and innovation program and EFPIA (www.imi.europe.eu).

## Contributions

P.W., M.V., L.S jointly organised the ACROBAT challenge. M.R., P.R. jointly supervised the ACROBAT challenge organisation. P.W., M.R., M.V. and P.R. jointly conceptualised the ACROBAT challenge. P.W. selected and verified the data set. P.W., M.V., LS. drafted this manuscript. P.W., L.S. created the figures in this manuscript. P.W. computed performance metrics and conducted statistical analyses. L.S., M.V. generated registration method overviews. P.W., K.K., M.V. processed the images. L.S., M.V. implemented the annotation infrastructure. M.V., L.S., P.W. generated the annotation instructions. C.C., C.B., S.K., A.K., D.R., Y.F., S.S.P., P.W., M.V., L.S., K.K., A.S., K.L.E. (in order of contribution) generated the landmark annotations. S.R. generated semantic annotations. M.R., J.H., P.R., A.L., L.L. acquired funding for this project. All authors contributed to editing the manuscript.

## Corresponding authors

Mattias Rantalainen (mattias.rantalainen@ki.se), Philippe Weitz (philippe.weitz@ki.se)

## Competing interests

M.R., J.H are co-founders and shareholders of Stratipath AB. K.K. is a co-founder and shareholder of Clinsight AB.

## Inclusion & Ethics

The study in whose terms the WSI data was generated has approval by the regional ethics review board (Etiksprövningsmyndigheten, Stockholm, Sweden, ref. 2017/2106-31 and amendments 2018/1462-32, 2019-02336). Due to the retrospective nature of the study, consent was not required.

# Supplementary Materials & Methods

**Supplementary Table 1.** Clinical characteristics of the 303 patients from which WSIs in the test set originate. Clinical characteristics for training and validation data were not extracted before anonymization.

| Clinical characteristic | Total number (percentage) |
|---|---|
| **Slide age** | |
| Mean | 1.64 years |
| Standard deviation | 1.93 years |
| **Cancer grade (NHG)** | |
| NHG1 | 63 (20.79%) |
| NHG2 | 139 (45.87%) |
| NHG3 | 94 (31.02%) |
| n.a. | 7 (2.31%) |
| **ER status** | |
| negative | 27 (8.91%)) |
| positive | 269 (88.78%) |
| n.a. | 7 (2.31%) |
| **PGR status** | |
| negative | 64 (21.12%) |
| positive | 232 (76.57%) |
| n.a. | 7 (2.31%) |
| **HER2 status** | |
| negative | 254 (83.50%) |
| positive | 42 (13.53%) |
| n.a. | 9 (2.97%) |
| **KI67 status** | |
| negative | 104 (34.32%) |
| positive | 170 (56.11%) |
| n.a. | 29 (9.57%) |

**Supplementary Table 2.** Percentages of areas and landmarks for each semantic segmentation annotation class in the test set. Disagreement indicates the percentage of landmarks for which the attribution between first and second annotator differs, potentially for landmarks close to borders of semantic segmentation regions.

|  | Tissue | Invasive cancer | Artefacts | DCIS | Non malignant changes | LCIS | Lympho-vascular invasion | Disagreement |
|---|---|---|---|---|---|---|---|---|
| Area | 61.88% | 32.65% | 2.44% | 1.68% | 1.20% | 0.13% | 0.01 % | n.a. |
| Landmarks | 48.99% | 33.55% | 1.30% | 7.95% | 3.46% | 1.01% | 0.09% | 3.64% |
| Factor | 0.79 | 1.03 | 0.53 | 4.73 | 2.88 | 7.77 | 9 | n.a. |

## Annotator distance

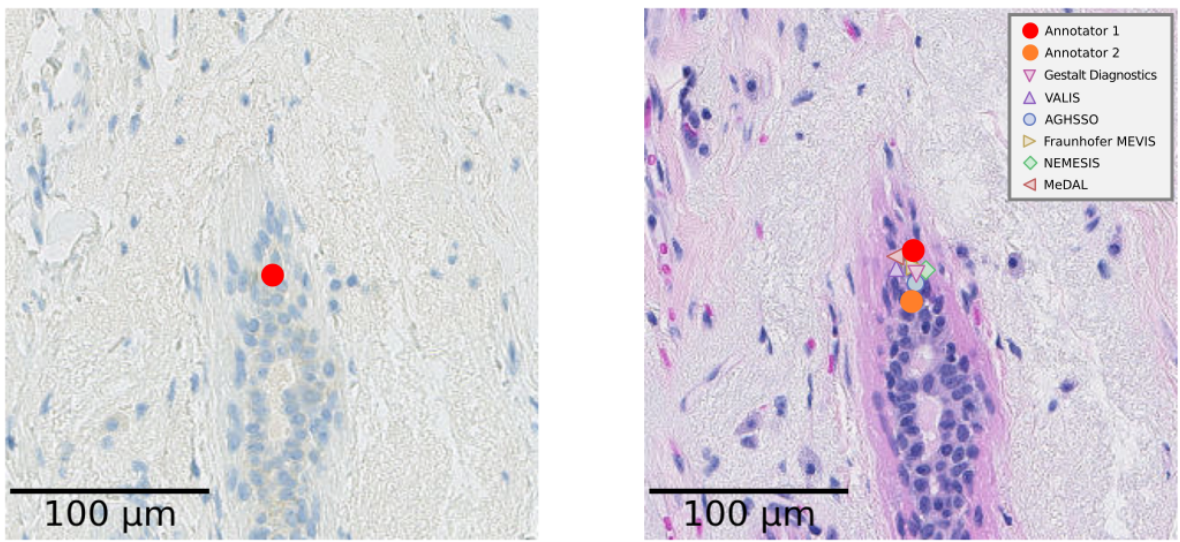

**Supplementary Figure 1**. Example of a landmark set in the IHC image by the first annotator and the corresponding selected locations by the first and second annotator in the corresponding H&E image. The submitted registered source landmarks by all teams are shown in the H&E image. If a registered landmark is between the two points selected by the two annotators, the TRE of this landmark is lower than the DBA.

# Effect of threshold on algorithm ranking

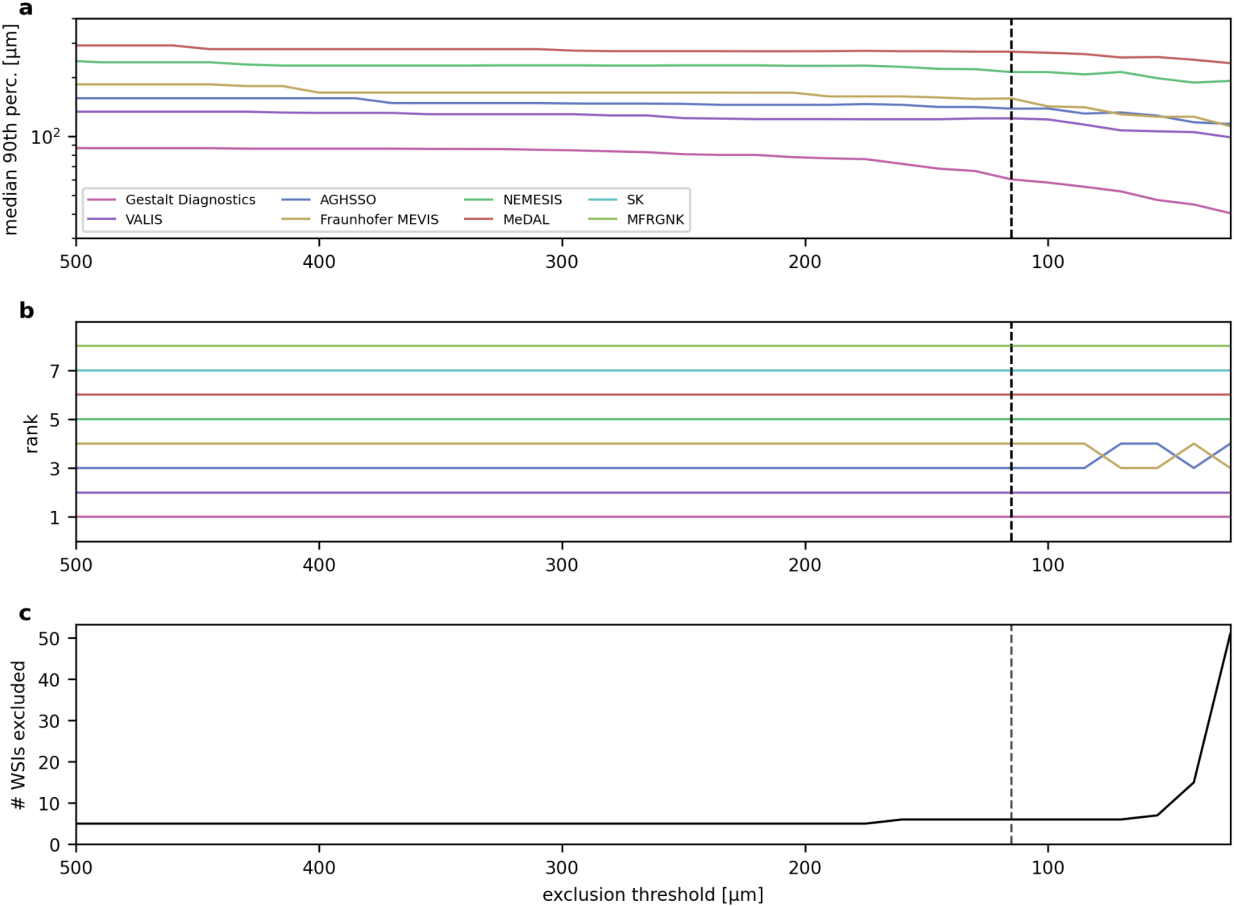

**Supplementary Figure 2.** Impact of the value of the threshold for annotator disagreement in μm on a) the median 90th percentile, b) the rank based on the median 90th percentile and c) the number of WSIs with fewer than 10 remaining landmarks, which were therefore excluded.

## Correlation between methods

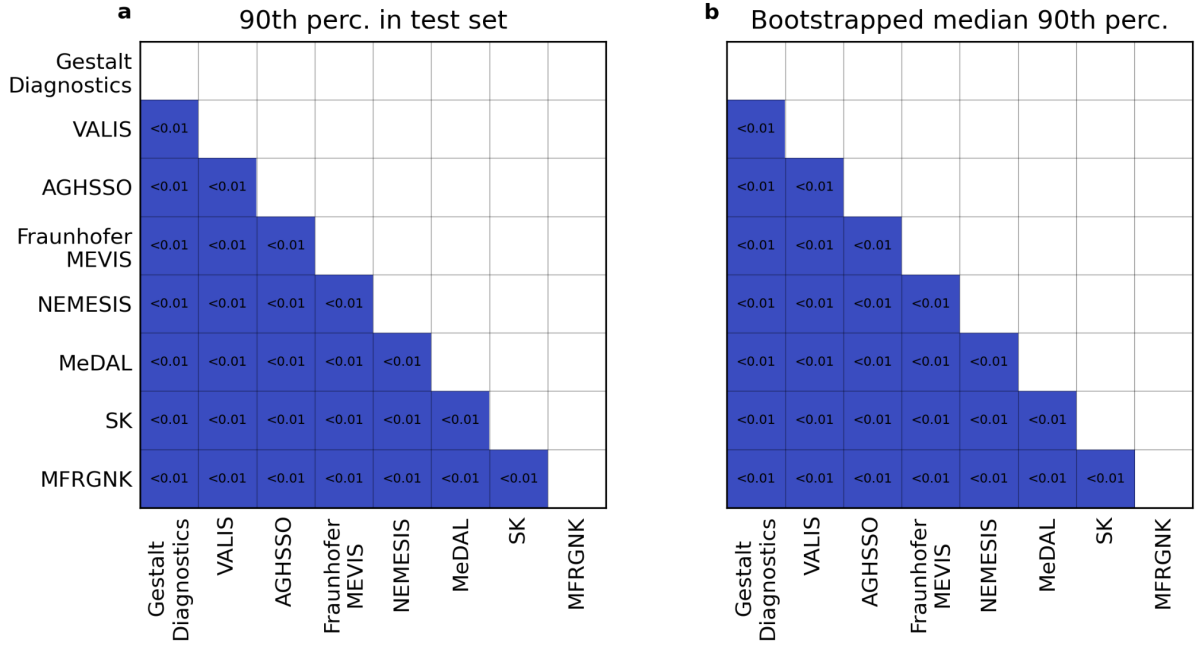

**Supplementary Figure 3.** Benjamini-Hochberg adjusted p-values for comparing the distributions metric in the test set with paired Wilcoxon signed rank tests. a) shows a comparison for the 90th percentiles, whereas b) shows a comparison for bootstrapped distributions of 90th percentiles.

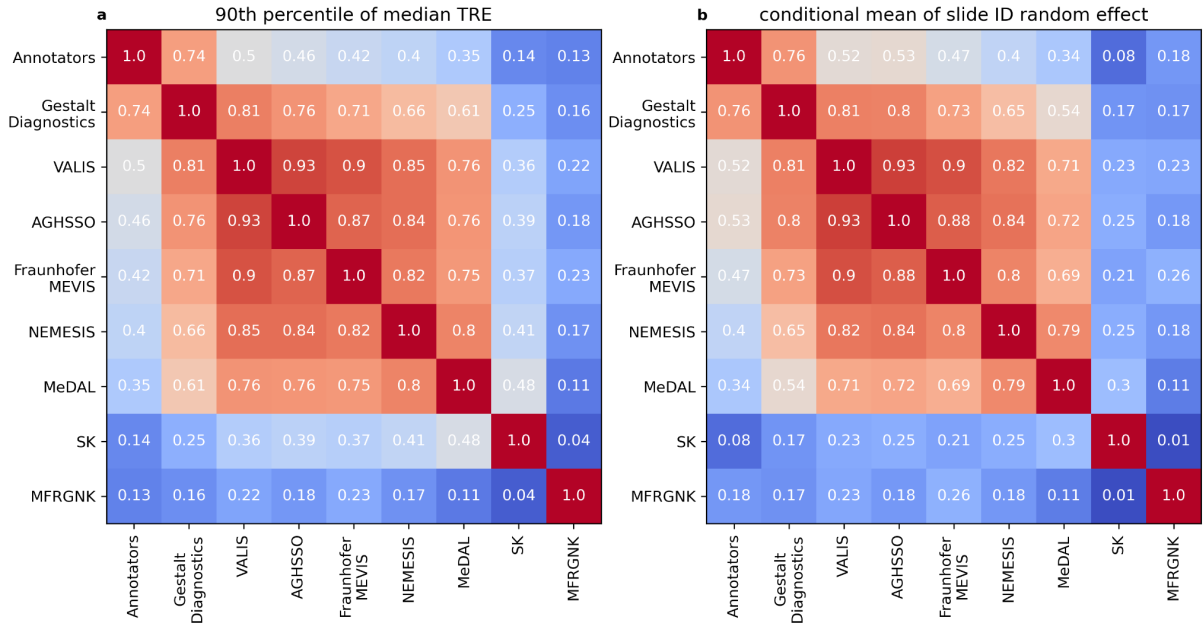

**Supplementary Figure 4.** Spearman correlations between the 90th percentiles of TREs in the test set in a), as well as the Spearman correlations of conditional means of the random effect that captures the slide ID in b).

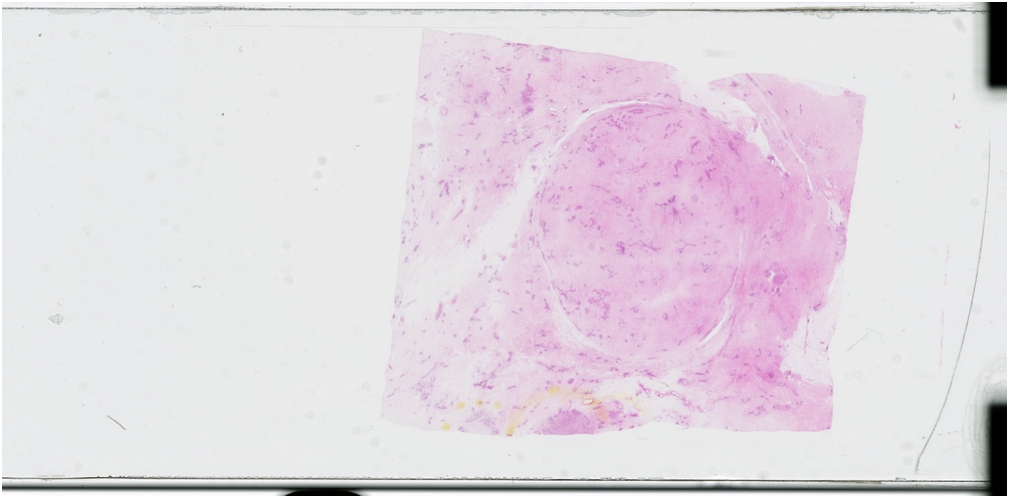
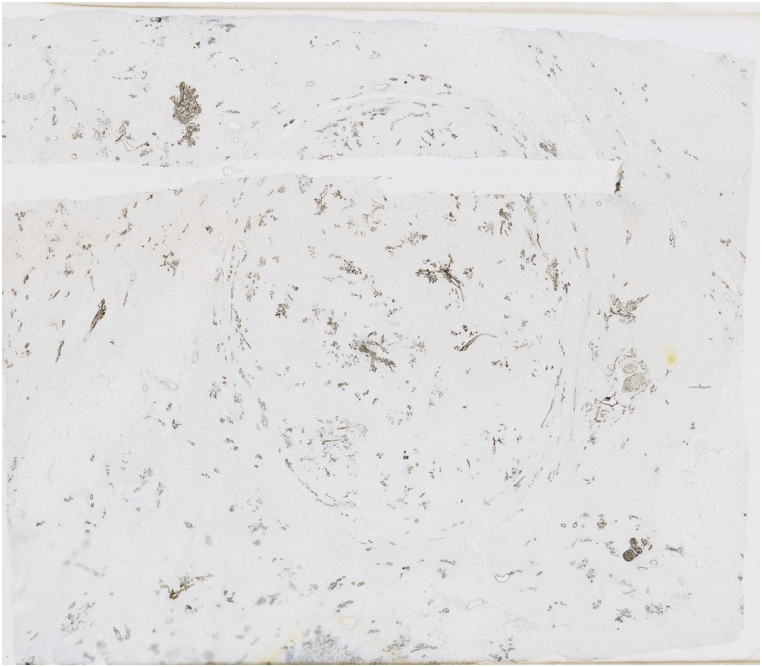

**Supplementary Figure 5.** The image pair with the worst mean 90th percentile across algorithms. The IHC WSI does not allow for the detection of a tissue outline due to the high degree of cropping.

**Supplementary Table 3.** Descriptions, ranges and references of LME model covariates.

|  | Range | Reference | Type | Explanation |
|---|---|---|---|---|
| Slide age | [0-5] | 0 years | Fixed effect | Time since diagnosis of the patient that the WSI of this landmark originates from as a continuous variable quantized in years, continuous. |
| Distance to center | [0-19] | 0 mm | Fixed effect | Distance to center of mass of the tissue mask in mm, continuous. |
| HER2 | {0, 1} | ER | Fixed effect | Indicator whether the IHC WSI is in the image pair of this landmark was stained for HER2. |
| PGR | {0, 1} | ER | Fixed effect | Indicator whether the IHC WSI in the image pair of this landmark was stained for PGR. |
| KI67 | {0, 1} | ER | Fixed effect | Indicator whether the IHC WSI in the image pair of this landmark was stained for KI67. |
| NMC | {0, 1} | Tissue | Fixed effect | Indicator whether this landmark is in an H&E image region that was annotated as non-malignant changes. |
| IC | {0, 1} | Tissue | Fixed effect | Indicator whether this landmark is in an H&E image region that was annotated as invasive cancer. |
| Artefact | {0, 1} | Tissue | Fixed effect | Indicator whether this landmark is in an H&E image region that was annotated as an artefact. |
| LCIS | {0, 1} | Tissue | Fixed effect | Indicator whether this landmark is in an H&E image region that was annotated as lobular carcinoma in situ. |
| DCIS | {0, 1} | Tissue | Fixed effect | Indicator whether this landmark is in an H&E image region that was annotated as ductal carcinoma in situ. |
| IC:NHG2 | {0, 1} | IC:NHG1 | Fixed effect | Interaction term that indicates whether a landmark belongs to the invasive cancer class and if so, whether this invasive cancer belongs to the NHG2 grade. |
| IC:NHG3 | {0, 1} | IC:NHG1 | Fixed effect | Interaction term that indicates whether a landmark belongs to the invasive cancer class and if so, whether this invasive cancer belongs to the NHG3 grade. |
| IC:BS:KI67 | {0, 1} | n.a. | Fixed effect | Interaction term that indicates whether a landmark is within an invasive cancer region and if so, if the IHC WSI in the image pair is stained for KI67 and if so, whether the IHC biomarker status is high. |
| IC:BS:HER2 | {0, 1} | n.a. | Fixed effect | Interaction term that indicates whether a landmark is within an invasive cancer region and if so, if the IHC WSI in the image pair is stained for HER2 and if so, whether the IHC biomarker status is high. |
| IC:BS:PGR | {0, 1} | n.a. | Fixed effect | Interaction term that indicates whether a landmark is within an invasive cancer region and if so, if the IHC WSI in the image pair is stained for PGR and if so, whether the IHC biomarker status is high. |
| IC:BS:ER | {0, 1} | n.a. | Fixed effect | Interaction term that indicates whether a landmark is within an invasive cancer region and if so, if the IHC WSI in the image pair is stained for ER and if so, whether the IHC biomarker status is high. |
| Slide ID | [0, 278] | n.a. | Random effect | Indicator of the slide ID that the landmark belongs to. |
| Annotator combination | [0, 86] | n.a. | Random effect | Indicator of the combination of first and second annotator that generated the two ground truth locations in the H&E WSI of this landmark. |

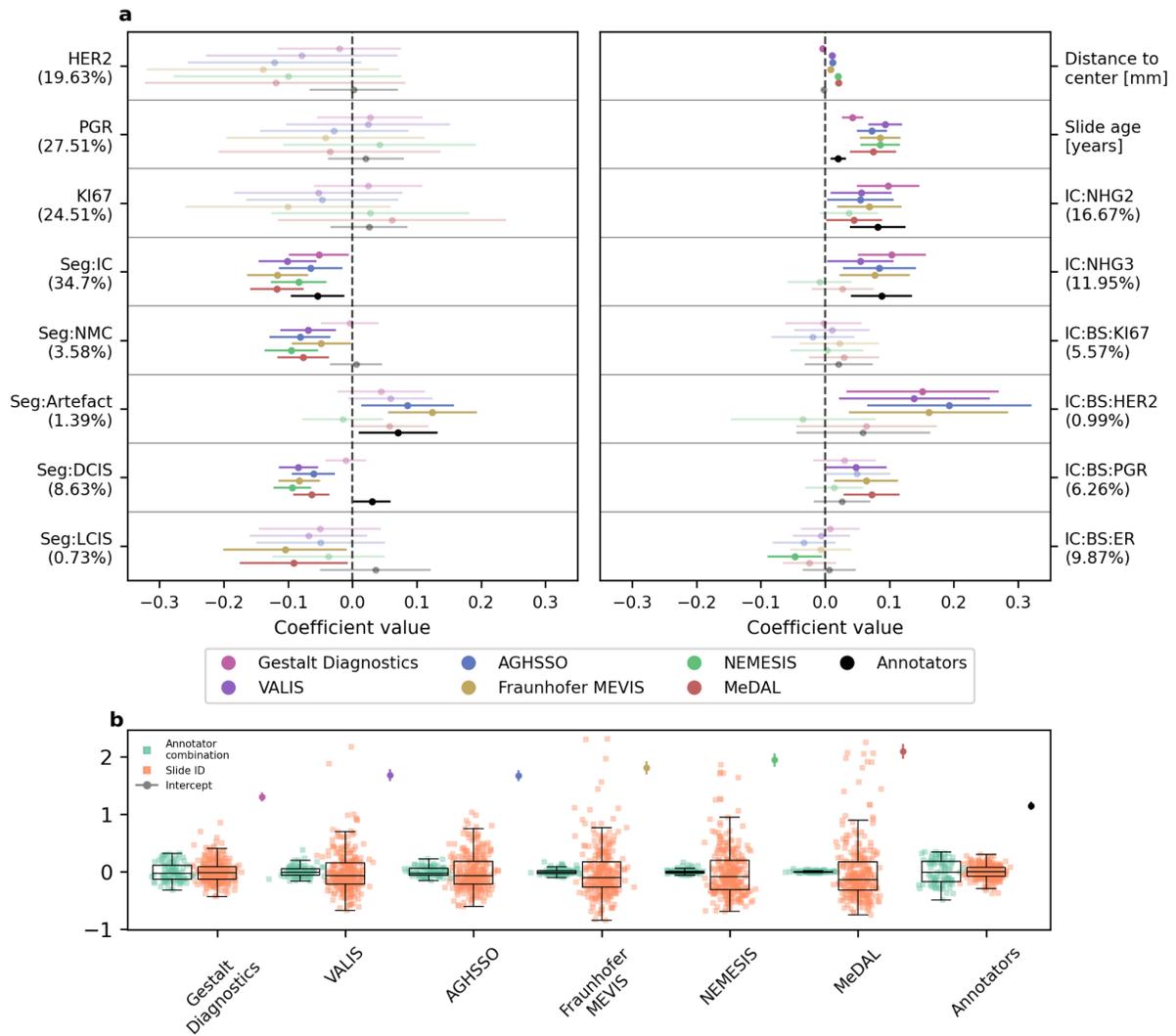

**Supplementary Figure 6. Coefficients and conditional means of random effects of the LME analysis.** a) shows the fixed effects coefficients with 95% confidence intervals for the LMEs fitted to the TREs of the six highest ranked teams, as well as for the LME fitted to the DBA. Transparency is increased if the confidence interval includes zero. For categorical fixed effects, there are indications of the percentage of landmarks that are part of the respective category. For continuous effects, the unit is indicated. Effects starting with *Seg* indicate landmark tissue classes, with normal tissue as the reference class. b) shows the distributions of the estimated conditional means of the random effects for the annotator combination and slide IDs, as well as the intercept for the respective LME model. Boxes include the lower to upper quartile of data. Whiskers extend 1.5 times the interquartile range from the box outlines or the minimum or maximum value.

**Supplementary Table 4.** Coefficients of the LME analysis for each fixed effect and team.

| | Gestalt Diagnostics | VALIS | AGHSSO | Fraunhofer MEVIS | NEMESIS | MeDAL | SK | MFRGNK | Annotators |
|---|---|---|---|---|---|---|---|---|---|
| Seg:LCIS | -0.051 [-0.146, 0.044] | -0.068 [-0.16, 0.023] | -0.049 [-0.15, 0.051] | -0.105 [-0.201, -0.008] | -0.037 [-0.125, 0.05] | -0.091 [-0.176, -0.007] | -0.029 [-0.09, 0.031] | 0.005 [-0.051, 0.061] | 0.036 [-0.05, 0.121] |
| Seg:DCIS | -0.01 [-0.042, 0.022] | -0.084 [-0.115, -0.054] | -0.06 [-0.094, -0.027] | -0.083 [-0.115, -0.05] | -0.094 [-0.123, -0.064] | -0.064 [-0.092, -0.035] | -0.052 [-0.072, -0.032] | -0.007 [-0.025, 0.012] | 0.03 [0.002, 0.059] |
| Seg:Artefact | 0.045 [-0.023, 0.113] | 0.059 [-0.007, 0.124] | 0.086 [0.014, 0.158] | 0.124 [0.055, 0.193] | -0.015 [-0.078, 0.047] | 0.058 [-0.003, 0.118] | -0.054 [-0.097, -0.01] | -0.132 [-0.172, -0.091] | 0.071 [0.009, 0.133] |
| Seg:NMC | -0.004 [-0.049, 0.041] | -0.069 [-0.112, -0.025] | -0.082 [-0.129, -0.034] | -0.049 [-0.094, -0.003] | -0.095 [-0.137, -0.054] | -0.077 [-0.117, -0.037] | -0.052 [-0.081, -0.023] | -0.073 [-0.099, -0.046] | 0.006 [-0.035, 0.046] |
| Seg:IC | -0.052 [-0.099, -0.006] | -0.101 [-0.146, -0.056] | -0.065 [-0.115, -0.016] | -0.116 [-0.164, -0.069] | -0.084 [-0.127, -0.041] | -0.117 [-0.159, -0.075] | -0.034 [-0.063, -0.004] | -0.007 [-0.035, 0.02] | -0.054 [-0.096, -0.013] |
| KI67 | 0.025 [-0.06, 0.109] | -0.053 [-0.184, 0.078] | -0.047 [-0.166, 0.071] | -0.101 [-0.26, 0.06] | 0.028 [-0.126, 0.182] | 0.061 [-0.116, 0.239] | 0.139 [-0.002, 0.278] | 0.069 [-0.031, 0.168] | 0.026 [-0.034, 0.085] |
| PGR | 0.028 [-0.055, 0.11] | 0.024 [-0.103, 0.152] | -0.028 [-0.143, 0.087] | -0.042 [-0.196, 0.112] | 0.042 [-0.108, 0.192] | -0.035 [-0.208, 0.137] | -0.033 [-0.169, 0.103] | 0.063 [-0.034, 0.159] | 0.021 [-0.038, 0.08] |
| HER2 | -0.02 [-0.117, 0.076] | -0.079 [-0.228, 0.07] | -0.121 [-0.257, 0.013] | -0.139 [-0.32, 0.041] | -0.1 [-0.278, 0.076] | -0.119 [-0.323, 0.083] | -0.026 [-0.186, 0.133] | -0.045 [-0.158, 0.067] | 0.002 [-0.067, 0.071] |
| IC:BS:ER | 0.008 [-0.038, 0.054] | -0.006 [-0.05, 0.039] | -0.033 [-0.082, 0.016] | -0.007 [-0.054, 0.04] | -0.047 [-0.09, -0.004] | -0.024 [-0.066, 0.017] | -0.042 [-0.071, -0.012] | -0.022 [-0.05, 0.005] | 0.006 [-0.035, 0.048] |
| IC:BS:PGR | 0.03 [-0.019, 0.079] | 0.048 [0.001, 0.095] | 0.049 [-0.002, 0.101] | 0.064 [0.014, 0.114] | 0.014 [-0.031, 0.059] | 0.072 [0.029, 0.116] | -0.052 [-0.083, -0.021] | 0.057 [0.028, 0.086] | 0.026 [-0.017, 0.07] |
| IC:BS:HER2 | 0.151 [0.033, 0.27] | 0.138 [0.021, 0.256] | 0.193 [0.066, 0.321] | 0.161 [0.037, 0.285] | -0.034 [-0.147, 0.079] | 0.064 [-0.044, 0.173] | 0.038 [-0.04, 0.117] | -0.041 [-0.114, 0.031] | 0.059 [-0.045, 0.164] |
| IC:BS:KI67 | -0.002 [-0.062, 0.057] | 0.011 [-0.048, 0.07] | -0.019 [-0.083, 0.045] | 0.022 [-0.04, 0.084] | 0.003 [-0.054, 0.059] | 0.029 [-0.025, 0.084] | -0.015 [-0.054, 0.024] | -0.064 [-0.101, -0.028] | 0.021 [-0.032, 0.074] |
| IC:NHG3 | 0.104 [0.051, | 0.055 [0.003, | 0.084 [0.027, | 0.077 [0.022, | -0.009 [-0.059, | 0.027 [-0.021, | -0.024 [-0.059, | 0.02 [-0.012, | 0.088 [0.04, |
|  | 0.157] | 0.107] | 0.141] | 0.132] | 0.041] | 0.076] | 0.01] | 0.053] | 0.135] |
| IC:NHG2 | 0.098 [0.05, 0.146] | 0.056 [0.009, 0.103] | 0.054 [0.003, 0.106] | 0.069 [0.019, 0.118] | 0.037 [-0.009, 0.082] | 0.045 [0.001, 0.089] | -0.029 [-0.06, 0.003] | 0.026 [-0.003, 0.056] | 0.082 [0.039, 0.125] |
| Slide age | 0.043 [0.026, 0.059] | 0.093 [0.067, 0.119] | 0.073 [0.049, 0.096] | 0.086 [0.054, 0.118] | 0.086 [0.055, 0.117] | 0.075 [0.039, 0.111] | 0.049 [0.021, 0.077] | 0.029 [0.009, 0.049] | 0.02 [0.009, 0.032] |
| Distance to center | -0.004 [-0.006, -0.001] | 0.01 [0.008, 0.013] | 0.011 [0.008, 0.014] | 0.008 [0.006, 0.011] | 0.02 [0.018, 0.023] | 0.021 [0.018, 0.023] | 0.039 [0.037, 0.041] | 0.013 [0.012, 0.015] | -0.002 [-0.005, 0.0] |
| Intercept | 1.302 [1.223, 1.379] | 1.682 [1.578, 1.785] | 1.675 [1.58, 1.768] | 1.807 [1.688, 1.925] | 1.947 [1.831, 2.063] | 2.097 [1.968, 2.225] | 2.817 [2.719, 2.915] | 3.848 [3.779, 3.916] | 1.151 [1.081, 1.218] |

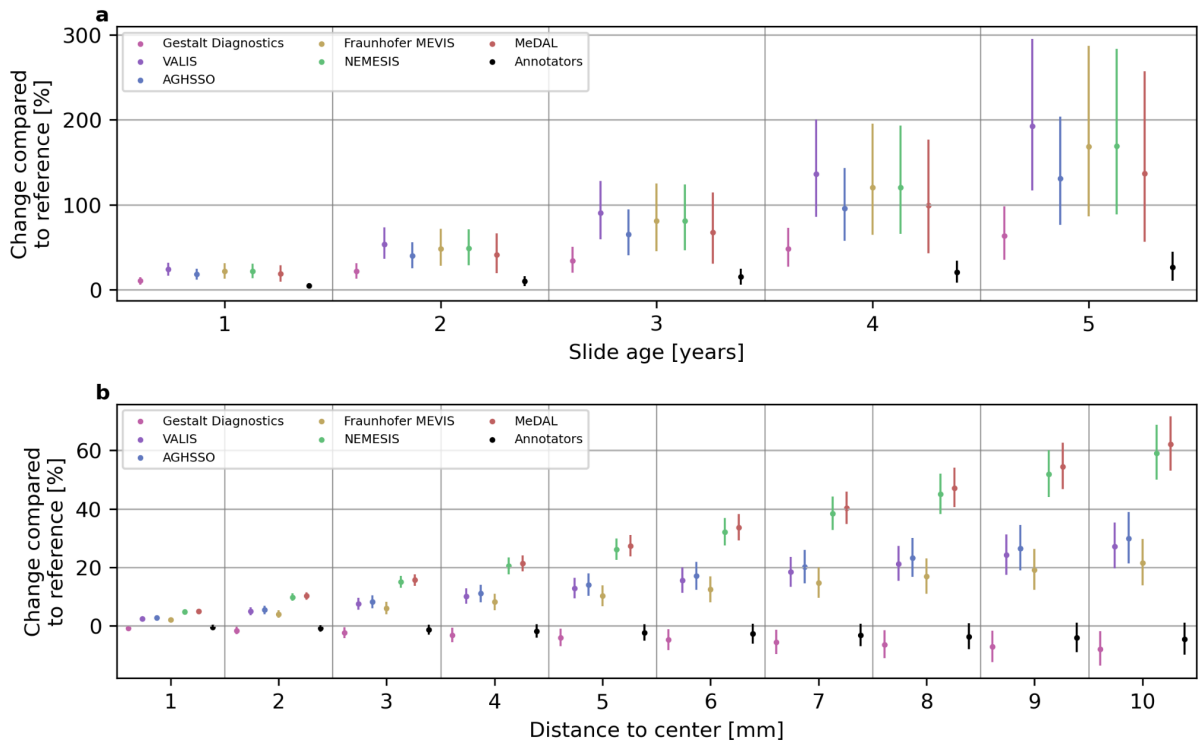

**Supplementary Figure 7.** Change in percentage compared to reference TRE for unit increases for the slide age in years in a) and the distance to the center of mass of tissue masks in mm in b).

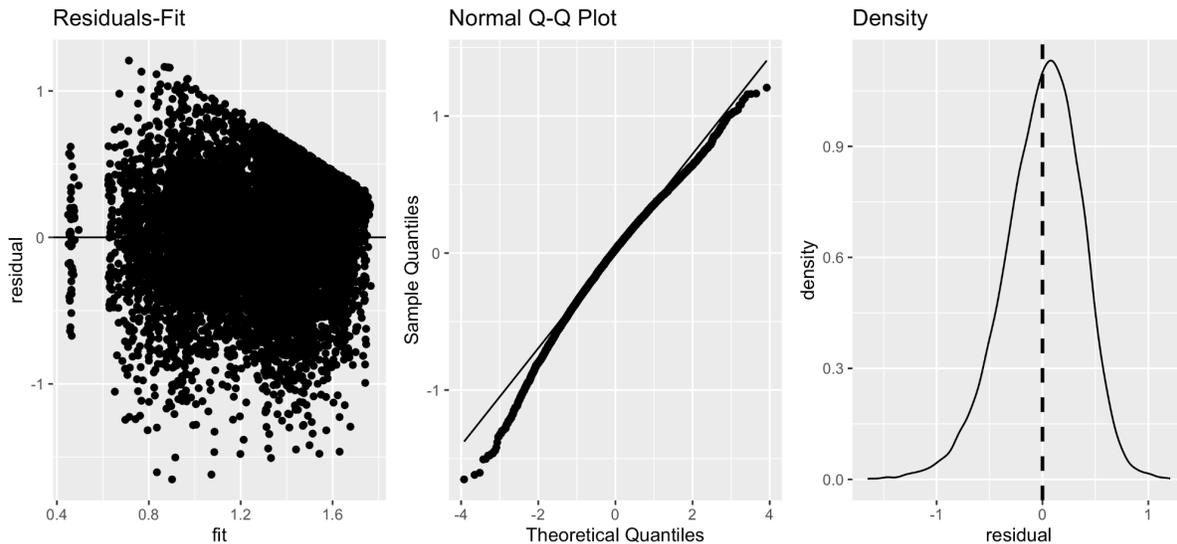

**Supplementary Figure 8.** Residuals-Fit, normal Q-Q plot and density of residuals for the LME model fitted to the DBA, with an exclusion threshold of 1 mm. Landmarks with a higher DBA than 1 mm are assumed to have been placed by accident.

## Supplementary Results - Scanner model, slide age and control tissue

The WSIs in this study were generated with three WSI scanners, the NanoZoomer S360 and two NanoZoomer XRs. The NanoZoomer XR scanners have different hardware revisions. Supplementary Figure 9a shows the percentage of image pairs for which the IHC WSI has control tissue. 9b shows the distribution of the years of diagnosis for each WSI scanned on the different scanners, whereas 9c depicts the percentage of IHC WSIs with control tissue for each year of diagnosis included. Supplementary Figure 9d shows the differences in mean TRE for each scanner for each of the six teams with the highest ranking. It is apparent that year of diagnosis, the presence of control tissue, the scanner model and the corresponding mean TREs are highly correlated. Since all WSIs that were digitized before 2015 were scanned on the XR2 scanner, it is not possible to disentangle the effect of the scanner and the year of diagnosis. We assume that the effect of aging and lower quality samples due to less automation in the slide preparation is the cause for the observed differences in performance between the scanners, but the available data does not allow a firm conclusion. Nevertheless, we did not include the scanner model or presence of control tissue in the LME analysis to avoid collinearity.

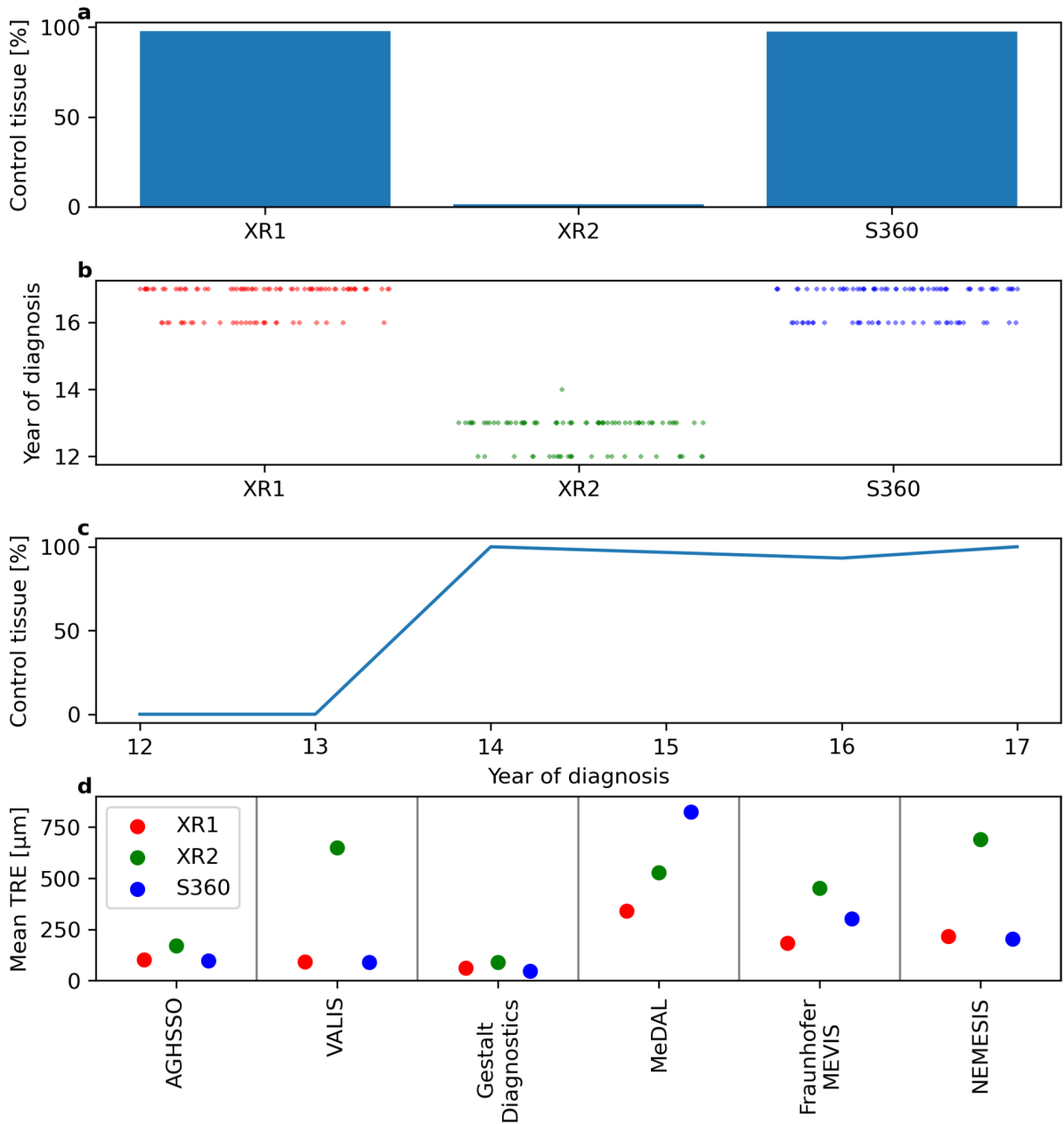

**Supplementary Figure 9.** Relationships between scanner model, year of diagnosis, control tissue and mean TRE for the six teams with the highest rankings. a) shows the percentage of image pairs for which the IHC WSI has control tissue for each WSI scanner. b) depicts the distribution of WSIs over years of diagnosis for each scanner, where each marker represents a WSI. c) shows the percentage of image pairs in which the IHC WSI has control tissue for each year of diagnosis included in the data set. d) depicts the mean TRE for each scanner for each of the six teams with the highest ranking.

## Overview of deployed algorithms

We asked the groups to answer, in their own words and writing styles, the following questions about pre-processing, pre-alignment, affine alignment, non-rigid registration, how ACROBAT data was used (and if any external data was used), what performance metrics were used during optimisation, main strengths and limitations of their methods.

| |
|---|
| Group: Gestalt Diagnostics |
| Citation: Proceedings of the MICCAI Workshop on Computational Pathology, PMLR 156:181-190, 2021 |
| Pre-processing:<br>Multiple patch-sizes are combined with contrast-normalising pre-processing to increase the number of matched key-points (RandomEqualize' and RandomSharpness from albumentations.ai) as a type of test time augmentation followed by non-maximum suppression. |
| Pre-alignment and Affine registration:<br>In the first step, the target image is rotated in 22.5-degree steps from 0 to 337.5, and the angle with the most matching key-points between the source/target image is taken as the initial rotation angle. Keypoints are extracted and matched from source and target image via SuperGlue, OpenGlue or LoFTR. Finally, calculating the affine matrix (OPENCV implementation for estimateAffine2D) with RANSAC filtering. |
| Non-rigid registration:<br>The non-rigid registration is approximated via a tree-based structure by reclusively applying affine transformations within the area of the source landmark until no new key-point pairs are found. The algorithm can be described as follows:<br>1) Perform key-point estimation and matching between the source and target image.<br>2) Perform Delaunay triangulation on the source key points and extract a new source and target patch pair for each triangle that contains a landmark to register.<br>3) Restart at step one and perform this reclusively until no new key-points are found. |
| Optimisation using ACROBAT data:<br>We used Weights and Biases Sweeps (docs.wandb.ai/guides/sweeps) to optimize the parameters for the key-point matching algorithms. Examples of these parameters are for superpoint and superglue: nms_radius, keypoint_threshold, max_keypoints, weights [outdoor or indoor], sinkhorn_iterations and match_threshold. |
| Use of external data:<br>Cases from the ANHIR challenge, the publication from Marzahl et al. (https://proceedings.mlr.press/v156/marzahl21a.html) and company internal datasets were used the test the robustness of the implementation and final parameterization. |
| Performance metric(s) during optimisation:<br>We tried to estimate the parameter configuration that finds the highest number of matching key-points which were not rejected by the RANSAC filtering. |

Main strength:
- Robust against tearing, folding, additional or missing tissue
- Imitates human behaviour by focusing on key areas
- Very accurate in regions of the slides where unique anatomical structures like vessels, glands or cell clusters are present.
- Optimized to register a limited number of landmarks or sparse cell types like mitotic figures for which this algorithm was initially developed and is applied in production.

Limitations:
- Inaccurate in areas with no unique structures like certain tissue types.
- It is significant time intensive to register WSI without focusing on a limited number of landmarks for guidance. Registering a WSI without landmarks needs multiple hours per image pair.
- To register a new landmark which is not preprocessed needs a rerun of the complete algorithm.

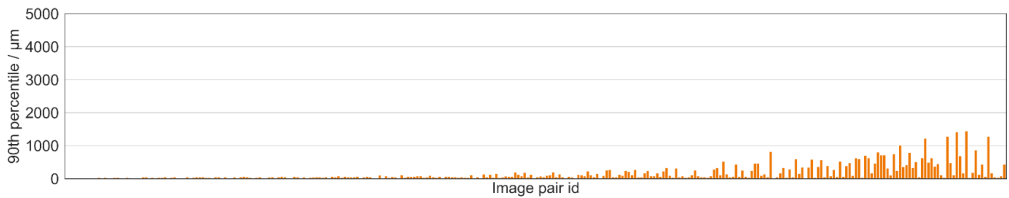

Code: Not available

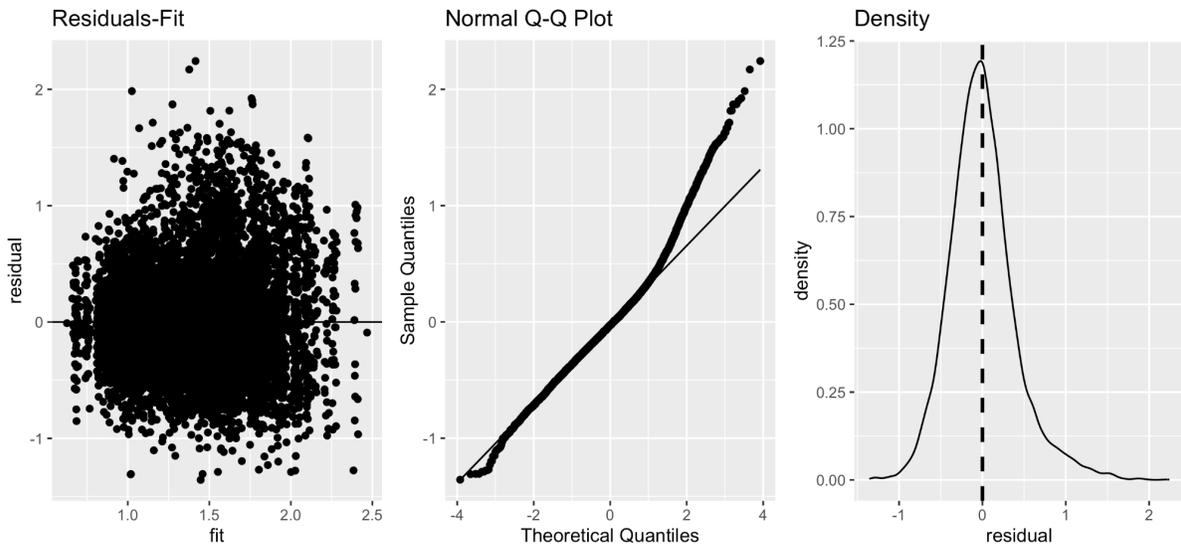

| |
|---|
| Group: VALIS |
| doi: 10.1101/2021.11.09.467917v1 |
| Pre-processing:<br>Each RGB image, I, was resized to have a maximum dimension of 850 pixels. Next, I was converted to polar CAM16-UCS colorspace and given a hue of 0 and colorfulness of 0.2. This image was then converted back to RGB, then to grayscale, and finally inverted to create image G. Pixels with luminosities (based on image I) greater than or equal to the 90th percentile of all luminosities were considered background, and their average value in G was subtracted from all pixels in G. G was then clipped to be between 0-1. Next, all dark regions touching the border in I were set to 0 in G. Adaptive histogram equalization was performed. Finally, the pairs of images were normalized to have a similar distribution of intensities. |
| Pre-alignment:<br>Features were detected using the BRISK algorithm, and then described using VGG. Brute force was used to match the features of each image pair. Poor matches were removed in two steps. First RANSAC was used to remove outliers. Next, we used Tukey's boxplot approach to identify outliers: a preliminary rigid transform, M', was found using RANSAC filtered matches, and then used to warp the feature points. The Euclidean distance (in pixel units) between the registered features was calculated, and only those between the lower and upper "outer fences" were kept. A similarity transform was then found using these filtered feature matches. |
| Affine registration:<br>No full affine transformation was estimated. |
| Non-rigid registration:<br>The bounding box of where the rigidly aligned images overlapped was scaled to have a maximum dimension of 850 pixels, and then used to slice the tissues from higher resolution, rigidly warped images. These higher resolution images were processed using the steps described above. Deep Flow was used to find the optical flow fields, F, used to align the moving image. The bounding box of the H&E tissue was then used to slice out even higher resolution tissues from the WSI, with the maximum dimension being ~25% of the WSI largest dimension. M and F were scaled and used to warp this higher resolution moving image, which was followed by a second non-rigid registration (also using Deep Flow) to improve alignment of the finer details. |
| Optimisation using ACROBAT data:<br>No answer |
| Performance metric(s) during optimisation:<br>Physical distance between matched features after outlier removal. But this metric was used more to test performance with/without different options, not for parameter tuning. |
| Main strength:<br>VALIS is stain agnostic and robust to noise, due to the preprocessing steps. These features are important because it makes VALIS generalizable. |
| Limitations:<br>While VALIS can potentially use the full resolution images for registration, it requires tiling and |

is very slow.

Comments:
While not relevant to the ACROBAT challenge, something readers might like to know about VALIS is that it is actually a groupwise registration method, and so can align multiple images simultaneously. Some may also like to know that VALIS can also register immunofluorescence images, and potentially brightfield to immunofluorescent.

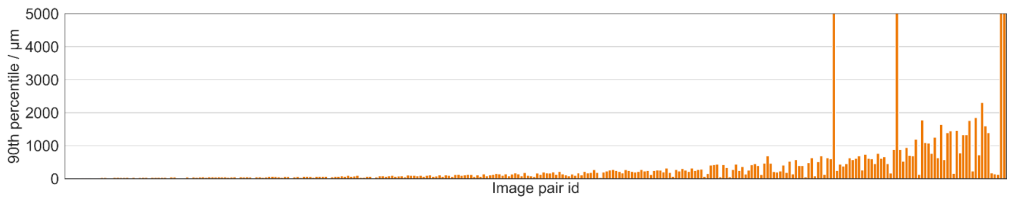

Code: https://github.com/MathOnco/valis

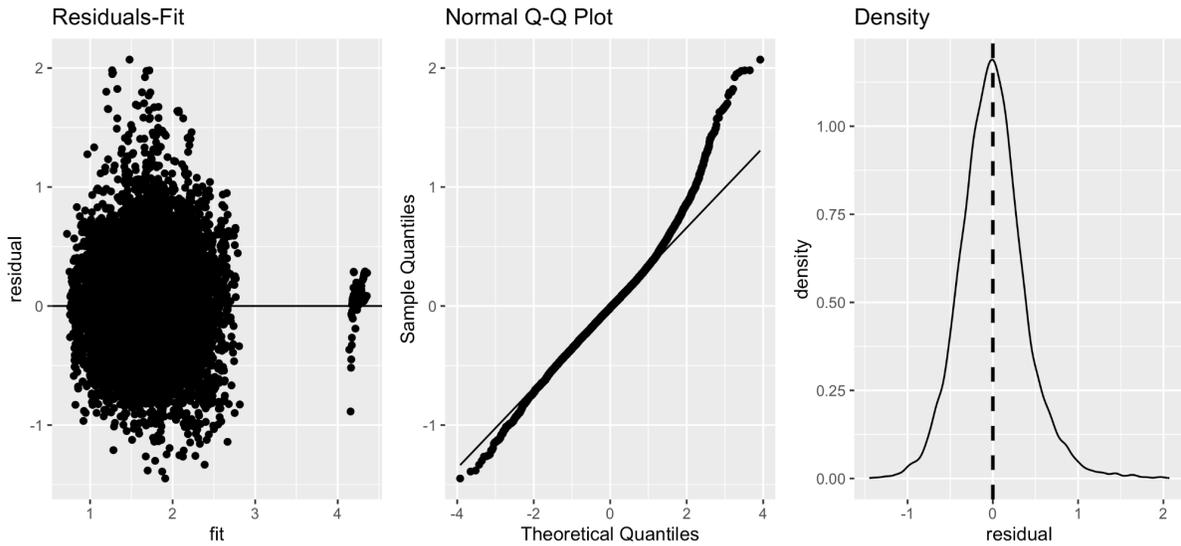

| Group: AGHSSO |
| --- |
| doi: https://doi.org/10.1016/j.cmpb.2020.105799 |
| Pre-processing:<br>The preprocessing consists of: (i) resampling the images to the resolution desired during the nonrigid registration step, (ii) padding the images to the same shape, (ii) converting the images to grayscale and inverting the intensities, (iv) equalizing the intensities using the CLAHE algorithm. |
| Pre-alignment:<br>The pre-alignment is based on a feature-based rigid/affine registration. The keypoints and features are extracted using SIFT and SuperPoint algorithms. The SuperPoint is applied without fine-tuning. The keypoints and descriptors are matched using RANSAC and SuperGlue methods. The process is repeated for several angles, and transformation types. The best transformation is chosen based on a sparse descriptor error. |
| Affine registration:<br>The affine refinement is based on iterative, intensity-based instance optimization using the Adam optimizer. The process is multi-level and the objective function is the local normalized cross-correlation (NCC). |
| Non-rigid registration:<br>The nonrigid registration is a multi-level instance optimization-based iterative procedure. The objective function is a weighted sum of the local NCC and the diffusive regularization. The method is run for a predefined number of iterations for each pyramid level with a constant learning rate. The regularization coefficient varies between levels. |
| Optimisation using ACROBAT data:<br>The training data of the ACROBAT dataset was not used since no model was trained. The registration parameters (number of levels, number of iterations per level, regularization coefficients, NCC window size, etc.) were tuned using the validation data and the challenge submission system. |
| Use of external data:<br>No external data was used to optimize the method performance. However, the method was verified on two other public datasets (ANHIR, HyReCo) and successfully registered all the cases from these datasets. |
| Performance metric(s) during optimisation:<br>Pre-alignment: sparse descriptor error (lowest mean squared error error between N most influential descriptors).<br>Affine registration: local normalized cross correlation,<br>Nonrigid registration: weighted sum of local normalized cross-correlation and diffusive regularization. |
| Main strength:<br>1)   Robust pre-alignment step working for almost all cases, even those with small overlap ratio between source and target tissue.<br>2)   Generalizability to different histology datasets without any fine-tuning since the method is |

iterative and is not learning based.
3) Simplicity and openness – the method is open source, the results are reproducible, the algorithm is easy to be reimplemented.

Limitations:
1) The pre-alignment step is overcomplicated. It should be simplified to further increase the computational efficiency and robustness.
2) The affine/nonrigid registration is iterative, therefore is not suited to real-time registration. The registration may take up to several minutes when using high resolution (e.g. > $5000^2$) or saturate the GPU memory (e.g. > $12000^2$).

Comments:
The method will be included and extended in the DeeperHistReg framework, to be released soon.

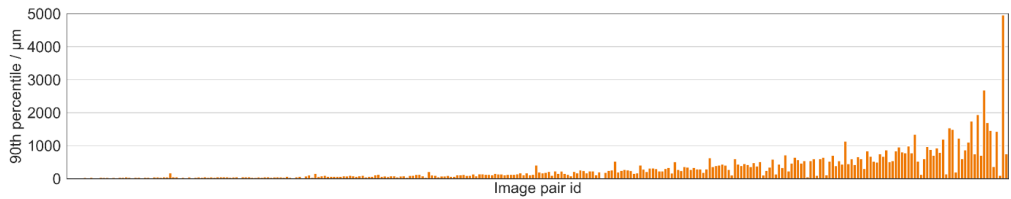

Code: https://github.com/MWod/DeeperHistReg-ACROBAT-submission

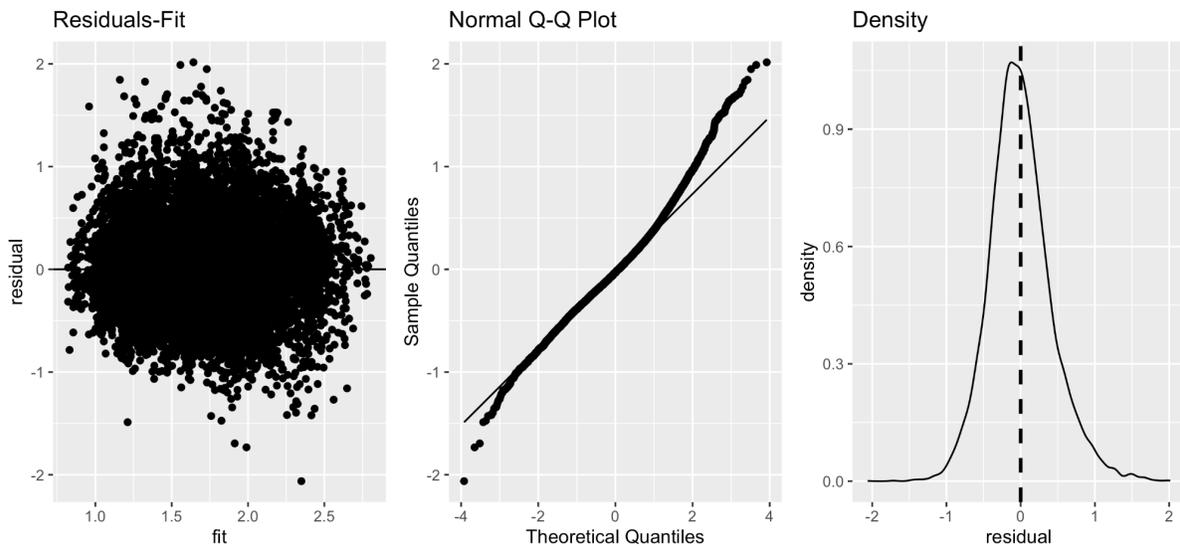

| Group: Fraunhofer MEVIS |
|---|
| doi:, preprint: https://doi.org/10.48550/arXiv.2106.13150 |

Pre-processing:
All images are converted from color to grayscale and inverted to obtain a black background while loading from disk. Normalized Gradient Fields (NGF) distance measure[1] is used throughout the registration steps. NGF is defined as

$$NGF(R,T,y) = \frac{h^2}{2}\sum_{i=1}^{N} 1 - \left(\frac{\langle \nabla T(y(\mathbf{x}_i)), \nabla R(\mathbf{x}_i)\rangle_\varepsilon}{||\nabla T(y(\mathbf{x}_i))||_\varepsilon \, ||\nabla R(\mathbf{x}_i)||_\varepsilon}\right)^2$$

with

$$\langle \mathbf{x},\mathbf{y}\rangle_\varepsilon = \mathbf{x}^T\mathbf{y} + \varepsilon^2, \; ||\mathbf{x}||_\varepsilon := \sqrt{\langle \mathbf{x},\mathbf{x}\rangle_\varepsilon},$$

and the edge parameter ε, which controls the sensitivity to edges in contrast to noise.

The pre-processing then consists of a foreground segmentation and a tissue selection step. We trained a segmentation network[2] (FCNN_C with a spacing of 8μm ≈ 1.25x magnification) on seven slides of public slide data[2], on six additional slides from the ACROBAT training data, and one internal image.

In a second step, the largest interconnected tissue parts are selected from each of the resulting tissue masks. We assume that there are a multitude of smaller (control) tissue parts that do not necessarily have a correspondence in the other slide of the pair.

To select the relevant tissue parts, the separate masks in each image are sorted by their surface area and the differences between the areas are computed. We then identify the largest gap in area sizes that divides the tissue parts in a large and a small group. Only the tissue parts in the large-area group are kept for further processing. Additional tissue parts are included to obtain the same number of tissue parts in both images of a pair.

Pre-alignment:
Automatic Rotation Alignment (ARA) first determines the center of mass[3] of both images, using the gray values of the pixels as the weights. The relevant, corresponding tissue parts are masked according to the segmentation above.

Let $(t_1,t_2)$ be the vector pointing from the center of mass of the reference image to the center of mass of the template image, and let $\Phi_k=2\pi(k-1)/(N_{rotations}-1)$, $k=1,...,N_{rotations}$ be equidistant rotation angles sampling the interval $[0,2\pi)$. For each angle, a rigid registration is computed with initial guess $(\varphi,t_1,t_2)$. Among all $N_{rotations}$ rigid registration results, the minimizer $(\Phi^*,t_1^*,t_2^*)$ with the smallest image distance is selected as an initial guess for the subsequent affine registration.

Affine registration:
In a second step, again an NGF-based image registration is computed based on the original, unmasked images. To allow for additional degrees of freedom, the registration is optimized with respect to an affine transformation $y_{affine}$ and based on a finer image resolution. The resulting transformation is then used as initial guess for a subsequent deformable registration. A Gauß-Newton optimization scheme[4] is used for the affine and rigid registration steps.

Non-rigid registration:
We use curvature regularization, which penalizes second-order derivatives of the displacement[5] and which has been shown to work very well in combination with the NGF distance measure[6,7]. As with the NGF distance, we evaluate the displacements in the pixel centers $x_1,...,x_m$ with uniform grid spacing $h$ and use finite differences to approximate the derivatives. Thus, the discretized curvature regularizer is defined as:

$$CURV(y) = \frac{h^2}{2} \sum_{i=1}^{m} |\Delta^h u_1(\mathbf{x}_i)|^2 + |\Delta^h u_2(\mathbf{x}_i)|^2$$

where $\Delta^h$ is the common 5-point finite difference approximation of the 2D Laplacian $\Delta=\delta_{xx}+\delta_{yy}$ with Neumann boundary conditions. In summary, for deformable registration, we minimize the objective function

$$J(R,T,y) := NGF(R,T,y) + \alpha CURV(y) \to min,$$

with respect to the deformation $y$. An L-BFGS optimization scheme[4] is used for the affine and rigid registration steps.

Optimisation using ACROBAT data:
We trained the tissue segmentation using ACROBAT and other data. A parameter study on ACROBAT data produced no notable differences in the parameter set (such as NGF edge parameter, smoothness parameter $\alpha$), compared to parameters used on the ANHIR data.

Use of external data:
The segmentation in the pre-processing was trained on external data.

Performance metric(s) during optimisation:
Normalized gradient fields (NGF)

Main strength:
The method is very robust towards unseen data. No modifications in the core registration algorithms had to be made in order to transition from ANHIR or other data. The method is also very fast and can register two slides in below one minute (depending on image resolution).

Limitations:
If a different number of tissue samples are present on the two slides of a pair, a dedicated tissue matching algorithm would be needed. Furthermore, the method relies on smoothness of the deformation. If deformations such as tears or folds violate these assumptions, these cannot be modelled.

Other comments:
The method optimizes not only with respect to image similarity but also with respect to curvature energy of the deformation. The resulting smoothness of the deformation field is important when computing deformed images and also copes with situations where insufficient image information is available (such as missing tissue parts).

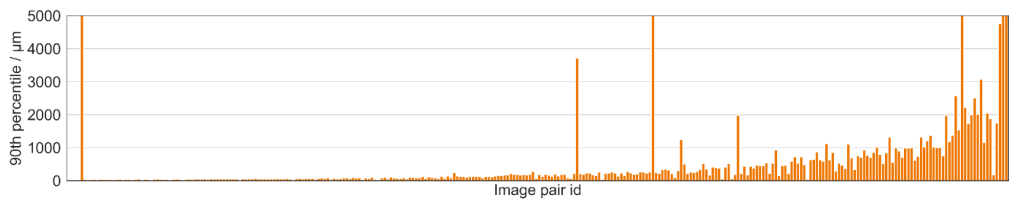

Code: Not available

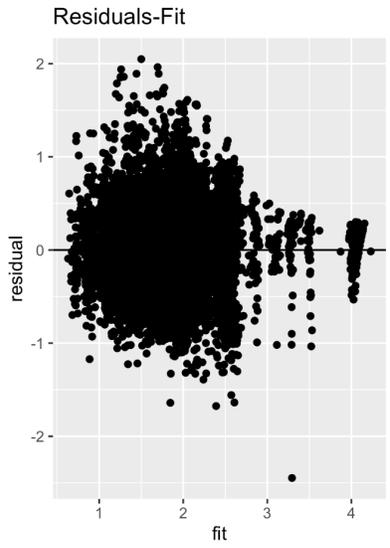
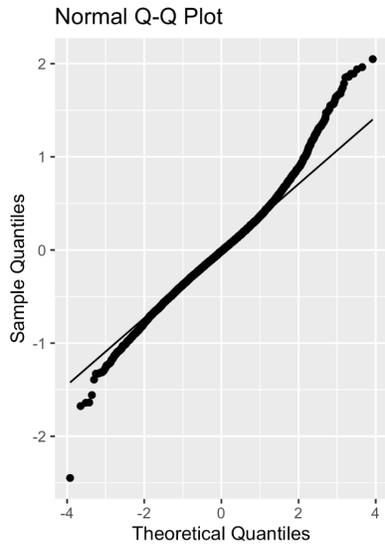
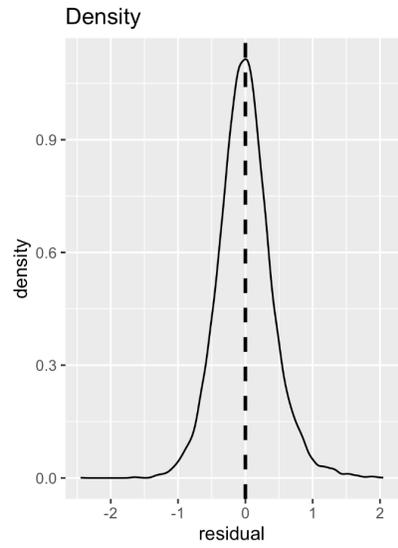

| Group: NEMESIS |
|---|
| Citation: openreview.net/forum?id=BP29eKzQBu3 |
| Pre-processing:<br>The first level of the pyramidal images wasdownscaled by a factor equal to 10. Before extracting the SIFT keypoints, the H&E RGB image was converted to the L*a*b* colorspace and the channel L was selected. The preprocessing consisted of four steps: i) a contrast limited adaptive histogram equalization (CLAHE), ii) Gaussian smoothing filtering with a $\sigma = 5$ μm , iii) taking the inverse of the image and iv) cropping the image to remove possible artefacts in the image edges (6 % of the image size for both image axes). Regarding the IHC preprocessing, after conversion from RGB to HSV colorspace, the aforementioned four preprocessing steps were applied to the V channel. Tissue was segmented in the whole slide images using the k-means algorithm (k = 3). The cluster with the highest average intensity was classified as the white background, hence, the tissue mask was defined as the union of the other two clusters. |
| Pre-alignment:<br>A conventional feature-based approach was used for the rigid pre-alignment. SIFT algorithm was used to extract relevant keypoints and Flann K-Nearest Neighbour (K=2) matching was used to match the keypoints. Matches with a Lowe's ratio below 0.95 were discarded as possible false matches. We used RANSAC to exclude outlier matches from the total set of matches. Matches with a registration error less than 50 pixels were defined as inliers. Additionally, for each iteration, the IHC tissue mask was rigidly transformed and the overlap, as Dice score, with the H&E tissue mask was computed. After 120 000 iterations, the best iteration was chosen by maximizing the following objective function: the squared sum of the number of inliers and the Dice score (after their min/max rescaling across the iterations). Also, a scale constraint was applied: iterations with a scale of the rigid transformation outside the range [0.95, 1.05] were discarded. |
| Affine registration:<br>No full affine transformation was estimated. |
| Non-rigid registration:<br>We optimized a MLP to find a transformation $\phi(x) = u(x) + x$ such that coordinate x in the IHC image corresponds to coordinate $\phi(x)$ in the H&E image. For further details please refer to the original paper . In this work, we used an MLP with three hidden layers, each of which contained 256 units with ReLU activation functions. Normalized cross correlation was used as loss function. An Adam optimizer was used with a learning rate of 1e−5 and 25 000 epochs. The preprocessed H&E L channel and the rigidly aligned IHC V channel at scale 1x were used as the fixed and moving image, respectively. During the MLP optimization, only coordinates belonging to the H&E tissue mask were sampled. In order to also include tissue edge information, the H&E tissue mask was dilated with a square structuring element of 51x51 pixels. |
| Optimisation using ACROBAT data:<br>The approach is unsupervised and the network is initialized and optimized separately for each image pair. However, some parameters were tuned by visual examination of the results (by observation of an overlay image between H&E and IHC image and the MLP training loss |

curve) on the validation set. The following parameters were tuned with this procedure: threshold on inlier distances in RANSAC, number of RANSAC iterations, MLP learning rate and number of training epochs.

Performance metric(s) during optimisation:
Normalised cross-correlation was used as loss function for the MLP optimization. The learning loss curves were analyzed to assess the performance of the network. Besides, we visually analyzed1) the results of matching SIFT keypoints before and after RANSAC, 2) the overlay between the H&E and the IHC images after the rigid pre-alignment and 3) the overlay between the H&E and the IHC images after the deformable registration  .

Main strength:
The use of implicit neural representations for image registration has advantages compared to traditional and convolutional neural network based methods. First, the deformation field is continuously defined over the image domain with a fixed number of parameters. Second, the method is not restricted to any particular image resolution and does not require the expensive task of computing discrete spatial gradients. This is particularly appealing in high-resolution whole-slide images. Third, one can control the ability of representing large or small local deformations by modifying the activation functions of the MLP.

Limitations:
Some limitations can be acknowledged: i) the rigid pre-alignment is failing in some cases and can produce unstable results if the number of RANSAC iterations is not high enough, but increasing it has a strong effect on the computational time; the introduction of the Dice score improves the performance of RANSAC but dramatically increases the computational cost. ii) in the current stage of the algorithm 1x scale images were used for the network optimization. However, work is in progress to optimize GPU memory usage and allow the use of the full scale image and therefore, increase the accuracy.

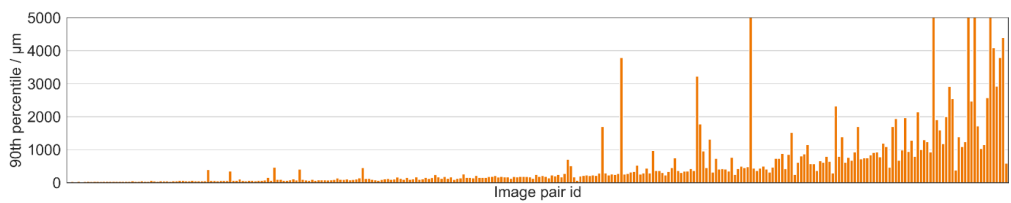

Code: https://github.com/MIAGroupUT/NEMESIS

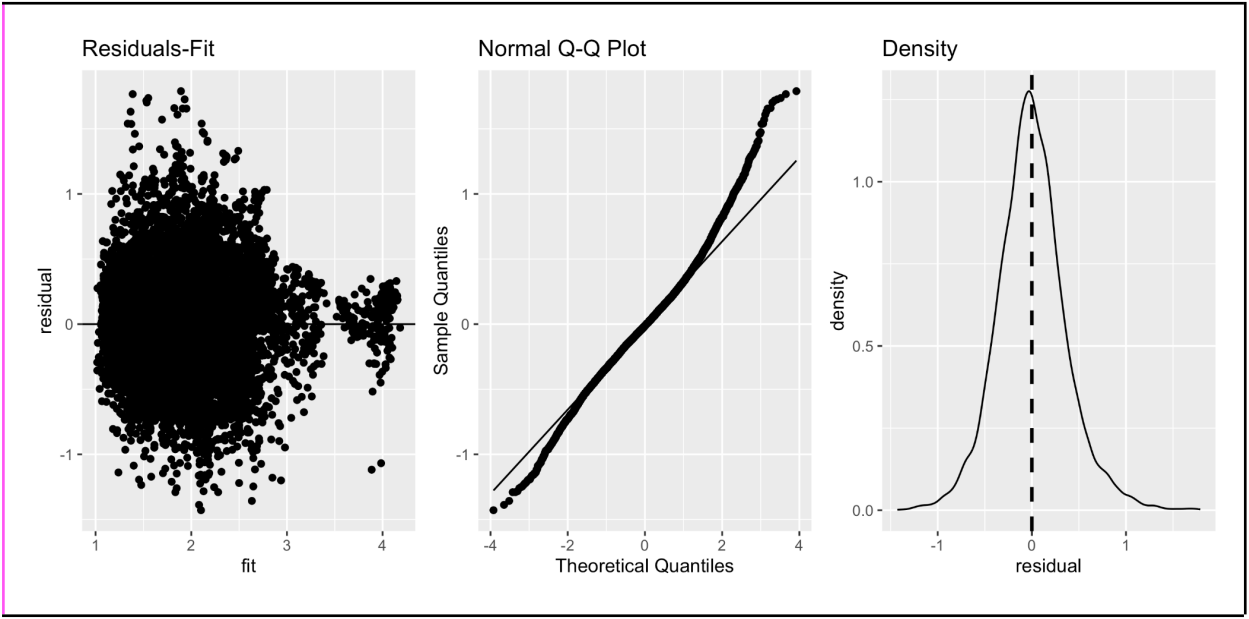

| Group: MeDAL |
|---|
| Pre-processing:<br>For both HE and IHC, we first identify foreground pixels with a thresholding operation. We perform this operation on downscaled images by a factor of 16. We convert these downscaled images to grayscale and apply higher and lower thresholds on pixel intensities. For HE, lower and higher thresholds used are 50 and 230, whereas, for IHC, thresholds are 100 and 240 respectively. We replace background pixels with 255 (maximum value) and invert the resulting images for further operations. At this stage, the tissue region of an image gets assigned relatively higher values than the background. |
| Pre-alignment:<br>We extract a bounding box containing tissue dense regions from HE image. We first calculate histograms across the x and y axis of the inverted HE image. We apply a threshold on these histograms to extract a tissue-dense region from the HE image. We use convolution operations for pre-alignment. We convolve the inverted bounding box of HE image (convolution kernel) with an inverted IHC image (input to convolution). We perform this operation on multiple rotations of HE ranging from 0 to 360 degrees with a stride of 1 degree. The argmax of convolution operation for all the rotations gives us three parameters – two positional (x and y) and one for rotation, which are the parameters for pre-alignment. We have performed convolution operations using pyTorch. |
| Non-rigid registration:<br>We use a radial basis function-based interpolation for non-rigid registration. First, we find displacement at particular key points in HE image. We obtain these key points using a foreground-background binary mask. We sample points from the boundary of this binary mask. The number of key points depends on the length of the tissue boundary in HE image. We extract patches from both HE and IHC images around a HE key point and corresponding (rigid) registered point from the IHC image. The size of HE patch at this stage is 256x256 whereas the size of IHC patch is 512x512. We perform a similar convolution operation as done in the rigid-registration stage at multiple key points to estimate local corrections at those key points. We then interpolate these local corrections using radial basis function-based interpolation operation. We have used the sklearn library for interpolation. We have described our method with diagrams here – https://github.com/abhijeetptl5/acrobat submission/blob/main/acrobat submission.pdf |
| Optimisation using ACROBAT data:<br>No optimisation |
| Performance metric(s) during optimisation:<br>No answer |
| Main strength:<br>Our method is easy to use and easy to explain. It requires installing a few stable python packages. eg. sklearn and pytorch. We can easily flag out the pair of WSIs for which registration performance is poor, based on few parameters in the algorithm. |
| Limitations:<br>Our algorithm is computationally heavy and the number of operations are heavily dependent on the size of WSI pairs. Also, we do not check the quality of WSI before performing |

registration in our pipeline, which can degrade performance of algorithm.

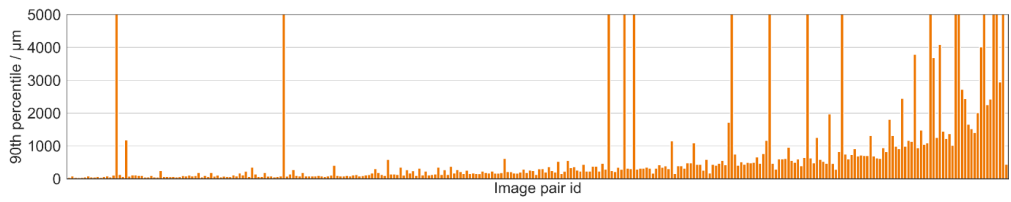

Code: https://github.com/abhijeetptl5/acrobat_submission

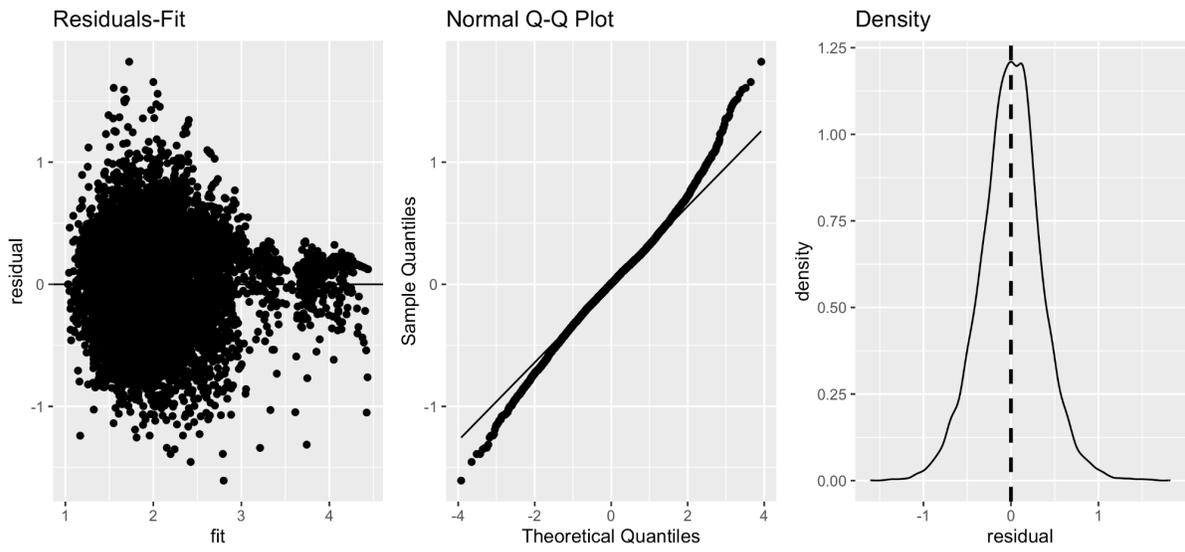

| Group: SK |
|---|
| Pre-processing:<br>Input images are downsampled by the ratio of 1/32. The input images (a fixed image and a moving image) are then converted to gray scale images. Some images have black area around the border of the image, and we remove this area by checking the pixel values in the border area. |
| Pre-alignment:<br>Our method estimates translation (in horizontal and vertical directions) and rotation (0 or 180 degree) between the fixed and moving images. We employ a template matching method to estimate translation and rotation. The matching criteria is normalized cross correlation. The template matching is performed by using the original shape moving image and the 180-degree rotated moving image. |
| Affine registration:<br>No answer |
| Non-rigid registration:<br>Our method performs non-rigid registration between the fixed image and the translated / rotated moving image. We employ VoxelMorph. In VoxelMorph, registration task is formulated as a function that maps an input image pair to a deformation field that aligns these images. The function is parameterized via a convolutional neural network (CNN), and the parameters of the neural network are optimized on a set of images. |
| Optimisation using ACROBAT data:<br>Tune the initial learning rates for the non-rigid transform and we used 0.001 as the initial learning rate at the final model. |
| Performance metric(s) during optimisation:<br>Normalized cross correlation for rigid transform (template matching). MSE loss for non-rigid transform (VoxelMorph). |
| Main strength:<br>Combination of rigid transform and non-rigid transform. |
| Limitations:<br>No answer |
| 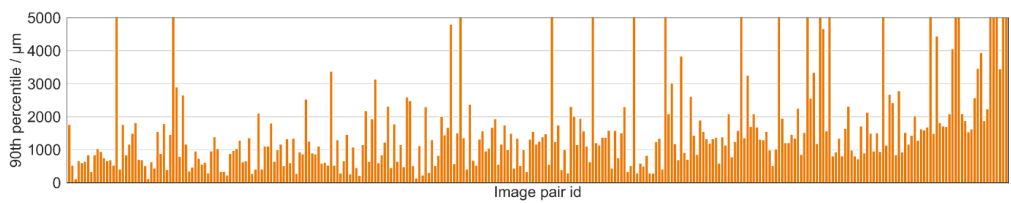 |
| Code: Not available |

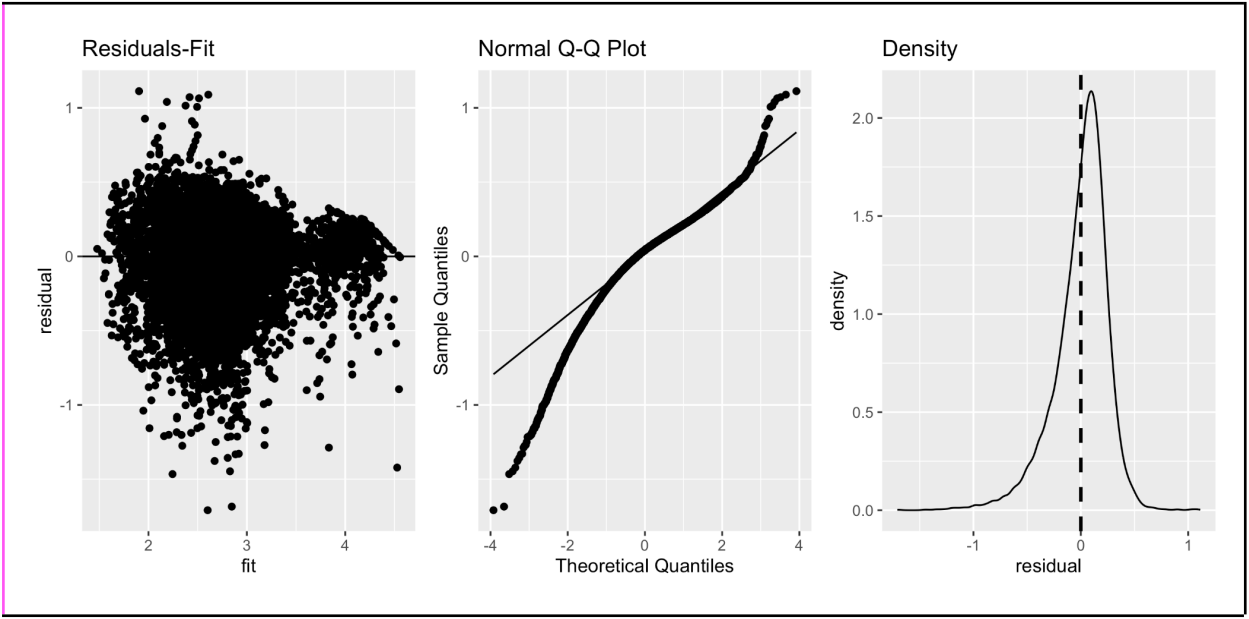

| Group: MFRGNK |
|---|
| Pre-processing:<br>IHC images provided with this challenge had on average 24k x 42k x 3 pixels, which were very large to process on our limited memory computers with GPUs. Thus, due to computation and time limitations, we reduced the dimension of every image to 0.05 of the original dimensions in each direction.<br>Then, we selected one H&E-stained tissue image as the reference color, and applied the Macenko stain normalizer method on all of the IHC images to make them more similar to the H&E-stained tissue image coloring. |
| Pre-alignment:<br>No answer |
| Affine registration:<br>No answer |
| Non-rigid registration:<br>For image registration between two IHC and H&E whole slide images(WSIs), We used the ORB detector to find 10k points of interest for each IHC and H&E WSIs, and we used the BFMatcher(Brute-force matcher using Hamming distance) on 10k descriptors. Next, we sorted the distance of these matched descriptions and selected the points in the top tertile (~30%).<br>Then, we found the homography matrix H, which is a nonsingular projective matrix that maps the points from a specific IHC image to the corresponding points in the H&E image. This homography matrix is found for each mapping of IHC to H&E images using the matched descriptions in the projective geometry.<br>In some cases, the estimation of the homography matrix is not accurate, meaning that it maps a point to another point that is not valid (negative position value or bigger than the image size). This issue happens due to the noise in images and the difficulty of mapping. In this case, we do the whole described method again in a higher resolution; for example, instead of resizing to 0.05, we resize the images to 0.1. We keep finding the homography matrix in higher resolutions until we find a feasible homography matrix. |
| Optimisation using ACROBAT data:<br>We tested our model with different numbers for points of interest and various resolutions for image registration. Increasing the number of PoI helped to achieve more accurate image registration. We know that doing image registration in higher resolution improves accuracy, but we decreased the image resolution due to computation and time limitations. Among different stain normalizer methods, we picked Macenko because it produced better results on the validation dataset. |
| Performance metric(s) during optimisation:<br>No answer |
| Main strength:<br>Our method is simple and can solve the problem at different resolutions, and it scales well with different computation resources. Moreover, our image registration is completely automated and works in real time. |
| Limitations: |

ORB detector is less robust on image transformations(rotation, translation and scaling). However, other methods, such as SIFT and SURF, can be used because they are more robust but require more computational time. We think with more time and computation resources, this model can be improved.

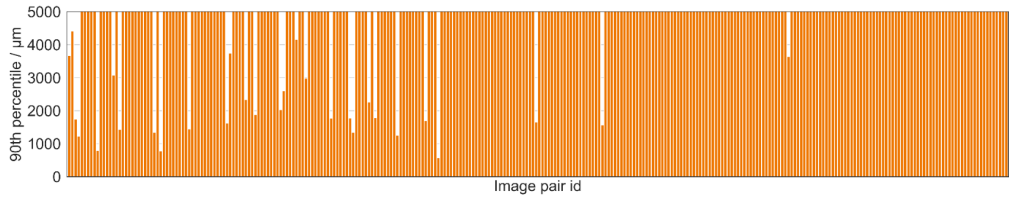

Code: https://github.com/mahtabfarrokh/ACROBAT-2022

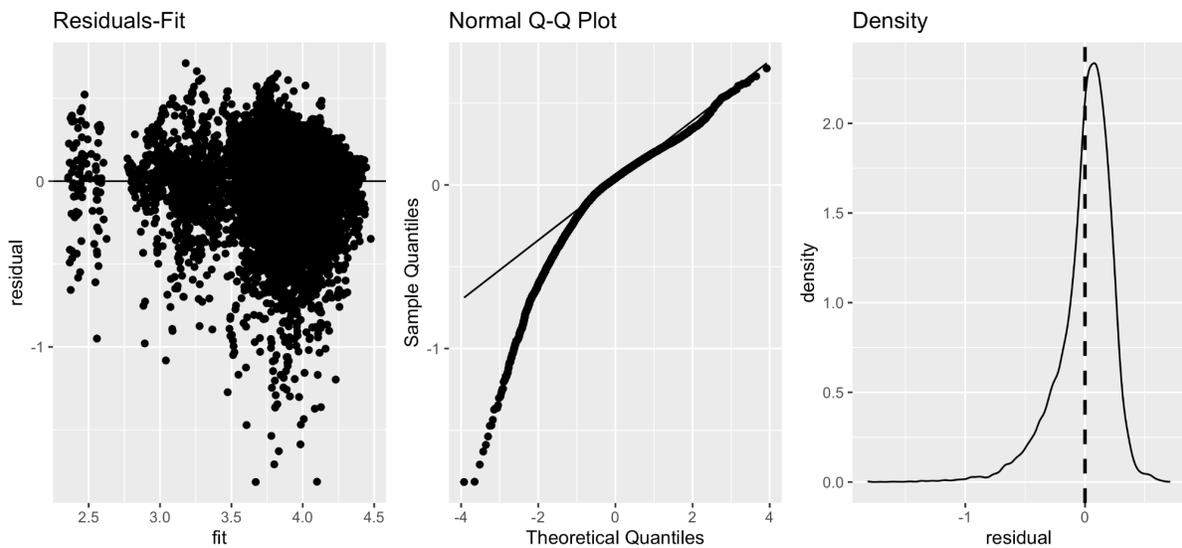

## Supplementary References